\documentclass[times,12pt]{article}

\usepackage{mdwlist}
\usepackage[utf8]{inputenc}

\usepackage{times,natbib}
\usepackage{balance}
\usepackage{epsfig}
\usepackage{graphicx}
\usepackage{amsmath, amsthm}
\usepackage{amssymb}
\usepackage{setspace}
\usepackage{multirow}
\usepackage[margin=1in]{geometry}
\usepackage{subfigure}
\RequirePackage[colorlinks,citecolor=blue,urlcolor=blue]{hyperref}
\usepackage{breakurl}
\usepackage{natbib,upgreek}
\usepackage{bbm}
\usepackage{multirow}
\usepackage{hyperref}
\usepackage{doi}
\usepackage{booktabs}
\usepackage{lineno}

\newcommand{\bfs}{{\bf s}}

\def\bSig\mathbf{\Sigma}

\begin{document}

\title{Detected changes in precipitation extremes at their native scales derived from in situ measurements}

\author{Mark D. Risser, Christopher J. Paciorek, Travis A. O'Brien,\\Michael F. Wehner, and William D. Collins}
\date{}
\maketitle

\begin{abstract}

\noindent The gridding of daily accumulated precipitation--especially extremes--from ground-based station observations is problematic due to the fractal nature of precipitation, and therefore estimates of long period return values and their changes based on such gridded daily data sets are generally underestimated. In this paper, we characterize high-resolution changes in observed extreme precipitation from 1950 to 2017 for the contiguous United States (CONUS) based on in situ measurements only. Our analysis utilizes spatial statistical methods that allow us to derive gridded estimates that do not smooth extreme daily measurements and are consistent with statistics from the original station data while increasing the resulting signal to noise ratio. Furthermore, we use a robust statistical technique to identify significant pointwise changes in the climatology of extreme precipitation while carefully controlling the rate of false positives. We present and discuss seasonal changes in the statistics of extreme precipitation: the largest and most spatially-coherent pointwise changes are in fall (SON), with approximately 33\% of CONUS exhibiting significant changes (in an absolute sense). Other seasons display very few meaningful pointwise changes (in either a relative or absolute sense), illustrating the difficulty in detecting pointwise changes in extreme precipitation based on in situ measurements. While our main result involves seasonal changes, we also present and discuss annual changes in the statistics of extreme precipitation. In this paper we only seek to detect changes over time and leave attribution of the underlying causes of these changes for future work.

\begin{center}
\end{center}

\end{abstract}



\onehalfspacing

\section{Introduction} \label{section1}


In spite of the increasing prevalence of high spatial resolution climate models and gridded daily products, it is important to utilize measurements from weather stations to characterize changes in the climatology of precipitation. This is especially true when considering changes in extreme precipitation, for which gridded data products may contain biases due to the fact that daily precipitation exhibits fractal scaling \citep[e.g.,][and numerous references therein]{Lovejoy2008,Maskey2016} and any smoothing technique will minimize extreme values. A recent thread of research documents these inherent problems with gridded data products and their use for characterizing climatological extremes \citep[see, e.g.,][]{King2013,Gervais2014,Timmermans2018,risser2018probabilistic}. As a result, changes in extreme precipitation calculated from gridded data products are potentially misleading. Without using gridded products, however, it is difficult to characterize spatially-coherent changes in extremes on the scales relevant for assessing impacts and designing infrastructure.  
 
There is a large existing literature on changes in extreme precipitation based on observations; these results fall into three general categories: (1)~changes derived from daily gridded precipitation products; (2)~changes based on station data, but only at the station locations; or (3)~changes based on station data, but aggregated into large areal averages. Across these different methods for presenting results, the literature explore changes in a variety of extremal metrics, for example the annual total precipitation divided by frequency of rainy days, the number of days that exceed a particular threshold or percentile, or return levels for a fixed duration.

The Intergovernmental Panel on Climate Change (IPCC) fifth assessment report \citep[][henceforth AR5]{ipcc_ar5} presents annual changes in extreme precipitation based on the HadEX2 dataset \citep{Donat2013}, which is a coarsely gridded (2.5$^\circ$ latitude by 3.75$^\circ$ longitude) dataset produced by calculating metrics of extreme precipitation and temperature at individual stations and then interpolating to a common grid. Figure 2.33 in the AR5 shows changes in the annual amount of precipitation from days greater than the 95th percentile as well as daily precipitation intensity. Both of these metrics exhibit increases over the central and eastern United States, some of which are \textit{very likely} significant, as well as slight decreases over the west and southwest United States. Overall, the AR5 finds that there are more regions with increases in the number of heavy precipitation events than decreases with the caveat that there are strong regional and seasonal differences, but in general it is difficult to ascribe a high degree of confidence in the magnitude and direction of the change.  

While the AR5 analyses are based upon a gridded product, \cite{Kunkel2003} explores changes in the United States using daily weather station measurements from the Global Historical Climatology Network \citep[GHCN;][]{Menne2012} from the late 19th through 20th centuries. {He} conclude{s} that overall the United States experienced an upward trend in extreme precipitation events since the 1920s and 1930s, but that extreme events in the late 1800s were as frequent as in the 1980s and 1990s. However, \cite{Kunkel2003} note{s} that natural variability in the climate system could be the cause of the observed increase over the last century. \cite{Balling2011} also use the GHCN data to describe annual changes in total precipitation over 1975-2010 {at individual weather stations}, identifying a general reduction in precipitation in the western USA and a general increase in the central and northeastern portions of the country. However, \cite{Balling2011} find that trends in precipitation display a large degree of spatial entropy or disorder wherein nearby stations show different or even opposite changes, and that the changes can differ for the same location depending on the metric and period of study {applied to the station data}. \cite{Kunkel2013} use GHCN measurements to {estimate} changes {during} 1948-2010 in the 20-year return value {at the locations of the GHCN stations}, finding that approximately three-fourths of all stations experience an increase in extreme precipitation but only 15\% {experience} a statistically significant increase. The central United States and North Atlantic show spatially-coherent {upward changes}, while the northwest United States shows less spatially-coherent {downward} changes.

Finally, many studies evaluate changes for large areal regions, where the changes are estimated from daily weather measurements that are first aggregated spatially. \cite{Groisman2005} use daily total precipitation data sets compiled at the National Climatic Data Center (including the GHCN) to evaluate changes in intense precipitation globally and find areas with both increases and decreases across the globe. Interestingly, they note that ``to obtain statistically significant estimates, the characteristics of heavy precipitation should be areally averaged over a spatially homogeneous region. Otherwise, noise at the spatial scale of daily weather systems masks changes and makes them very difficult to detect'' \citep{Groisman2005}. \cite{Janssen2014} explore trends in the GHCN data over 1900-2012 for seven large subregions of the contiguous United States (CONUS) {and obtain} an overall increase in extreme precipitation. Statistically significant increasing trends in extreme precipitation {at the 95\% confidence level} occur over the Midwest and Eastern regions, while most decreasing trends occur over Western regions, consistent with {the results of} \cite{Kunkel2003} and \cite{Groisman2005}. \cite{Easterling2017} evaluate seasonal changes in the 20-year return value based on the GHCN measurements over 1948 to 2015 and find major regional differences in the changes with the largest increases occurring in the northeastern United States. \cite{Easterling2017} identify the largest positive changes in fall and winter {when} all but one of  seven large CONUS subregions {exhibit} increases but do not indicate where and when a degree of confidence might be ascribed to the changes.

In summary, we reiterate that the literature contains a diverse set of data analyses focusing on observed changes in extreme precipitation over CONUS and that most of these come to common conclusions about the {sign} of the changes. Of the aforementioned studies, only \cite{Donat2013} present spatially-{continuous} trends based on irregularly-spaced weather station observations, however they do so on a very coarse grid via a distance-based weighting algorithm \citep{Donat2013}.  Among the remaining studies, many fail to ascribe a confidence statement to the results. Furthermore, many of these analyses only evaluate annual changes, which ignores important seasonal differences in extreme precipitation that should be accounted for when characterizing changes.

In this paper, we characterize changes in observed extreme precipitation for a fine $0.25^\circ$ spatial grid over the contiguous United States for 1950-2017 using station data only, utilizing spatial statistical methods to derive gridded changes based on irregularly observed measurements. Furthermore, we use a robust statistical technique to identify significant pointwise changes in the climatology of extreme precipitation while also carefully controlling the rate of false positives. The use of spatial statistics yields both an increased signal to noise ratio for the changes as well as insight into the physical behavior of changes in the climatology of extreme precipitation. 
While we conduct separate analyses for each season, we also derive annual changes in extreme precipitation based on the seasonal results.

To be clear, the results presented here only seek to detect the presence 
of changes in extreme precipitation and specifically do \textit{not} seek to characterize temporal trends or attribute the underlying causes of these changes. In characterizing smooth (linear) trends over time only, we do not intend to account for sources of interannual variability (e.g., the El Ni\~no Southern Oscillation) or decadal to multidecadal sources of variability (e.g., the Pacific Decadal Oscillation or the Atlantic Multidecadal Oscillation) that are known to influence precipitation extremes in the  contiguous United States. Given the relatively short time period considered in this analysis (1950 to present), it would be essentially impossible to account for these low frequency oscillations using observational data only because any changes we identify over the 68 years considered could be due to, e.g., the Atlantic Multidecadal Oscillation (which is roughly monotone over this period). However, given that our methodology provides a better characterization of precipitation extremes and allows for a high-resolution assessment of changes based on station data, we argue that our results are nonetheless an important addition to the literature on empirical changes in extreme precipitation. For future work, we plan to pursue a formal attribution of the underlying causes of any changes in extreme precipitation that may appear in the historical record.

The paper proceeds as follows: in Section \ref{section2}, we describe the weather station data set used in our analysis, and in Section \ref{StatMethods} we describe our statistical methodology. Results are presented in Section \ref{sec:results}, and Section \ref{sec:discussion} concludes the paper.

\section{Data} \label{section2}

\begin{figure}[!t]
\begin{center}
\includegraphics[trim={0 27 0 16mm}, clip, width = \textwidth]{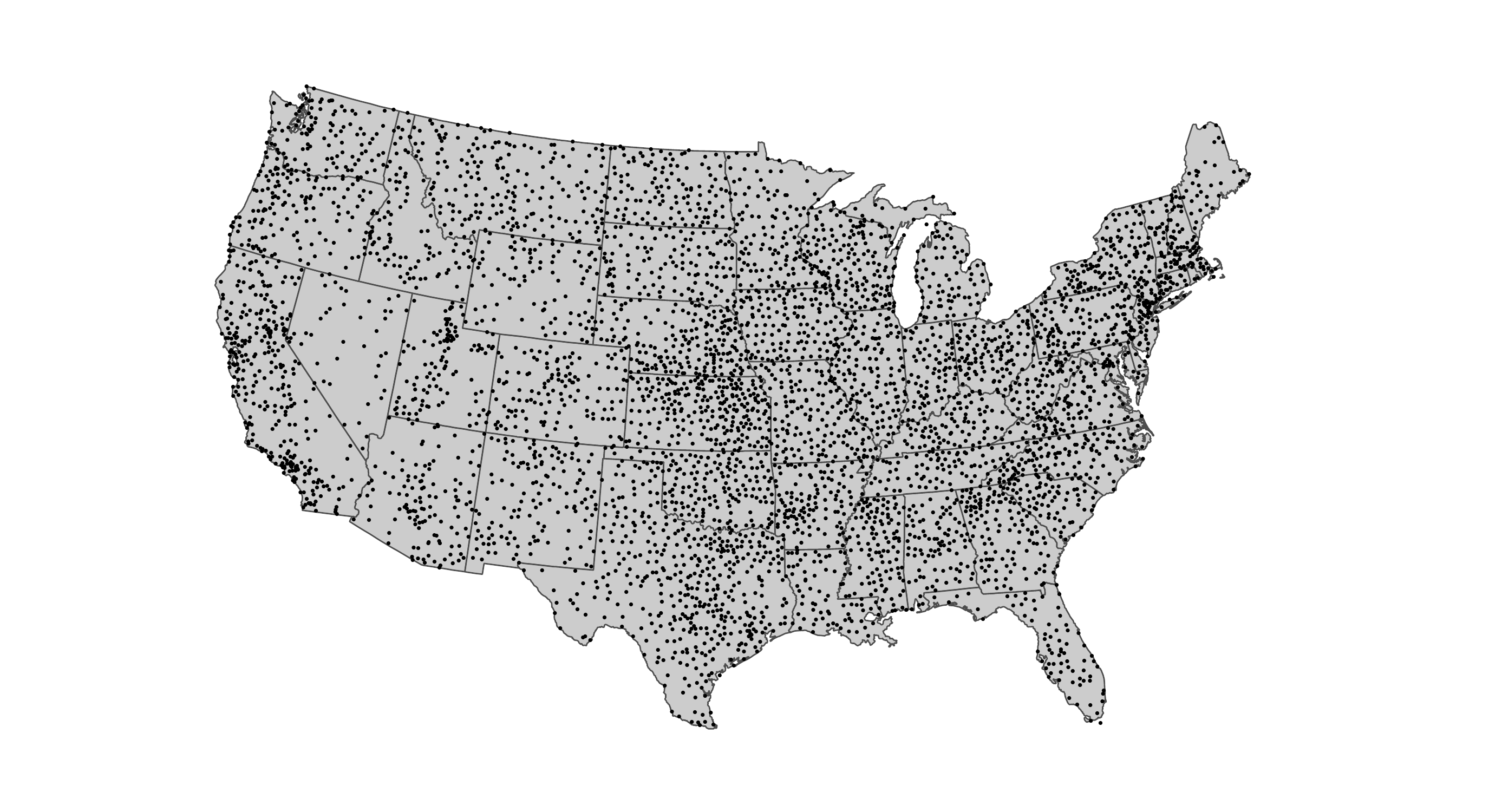}
\caption{The spatial distribution of the $n = 5202$ GHCN stations with a minimum of 66.7\% of daily precipitation measurements over December, 1949 to November, 2017.}
\label{GHCNstations}
\end{center}
\end{figure}

The data used for the following analysis consist of daily measurements of total precipitation (in millimeters) obtained from the Global Historical Climatology Network \citep[GHCN;][]{Menne2012} over the contiguous United States (CONUS). We refer the reader to \cite{risser2018probabilistic} for details on the quality control procedure.
{A}fter pre-processing the daily values, we select the subset of stations that {have} a minimum of 66.7\% nonmissing daily precipitation measurements {between} December~1,~1949 through November~30,~2017. This procedure {yields} a high-quality set of daily precipitation measurements spanning $T=68$ years for $n = 5202$ stations shown in Figure  \ref{GHCNstations} {down-selected from} the entire GHCN database of over twenty thousand weather stations. All subsequent analysis in this paper is based on seasonal maxima $\{ Y_t^{(j)}(\bfs) : t = 1950, \dots, 2017; j \in \{\text{DJF, MAM, JJA, SON} \}; \bfs \in \mathcal{S} \}$, where $\mathcal{S} = \{\bfs_1, \dots, \bfs_n\}$ denotes the $n=5202$ stations shown in Figure \ref{GHCNstations}. Note that the year index $t$ represents a ``season year'' such that, for example, the 1950 DJF is defined as December, 1949 to February, 1950. 

\section{Statistical methods} \label{StatMethods}

\begin{figure}[!t]
\begin{center}
\includegraphics[trim={0 0 0 0mm}, clip, width = \textwidth]{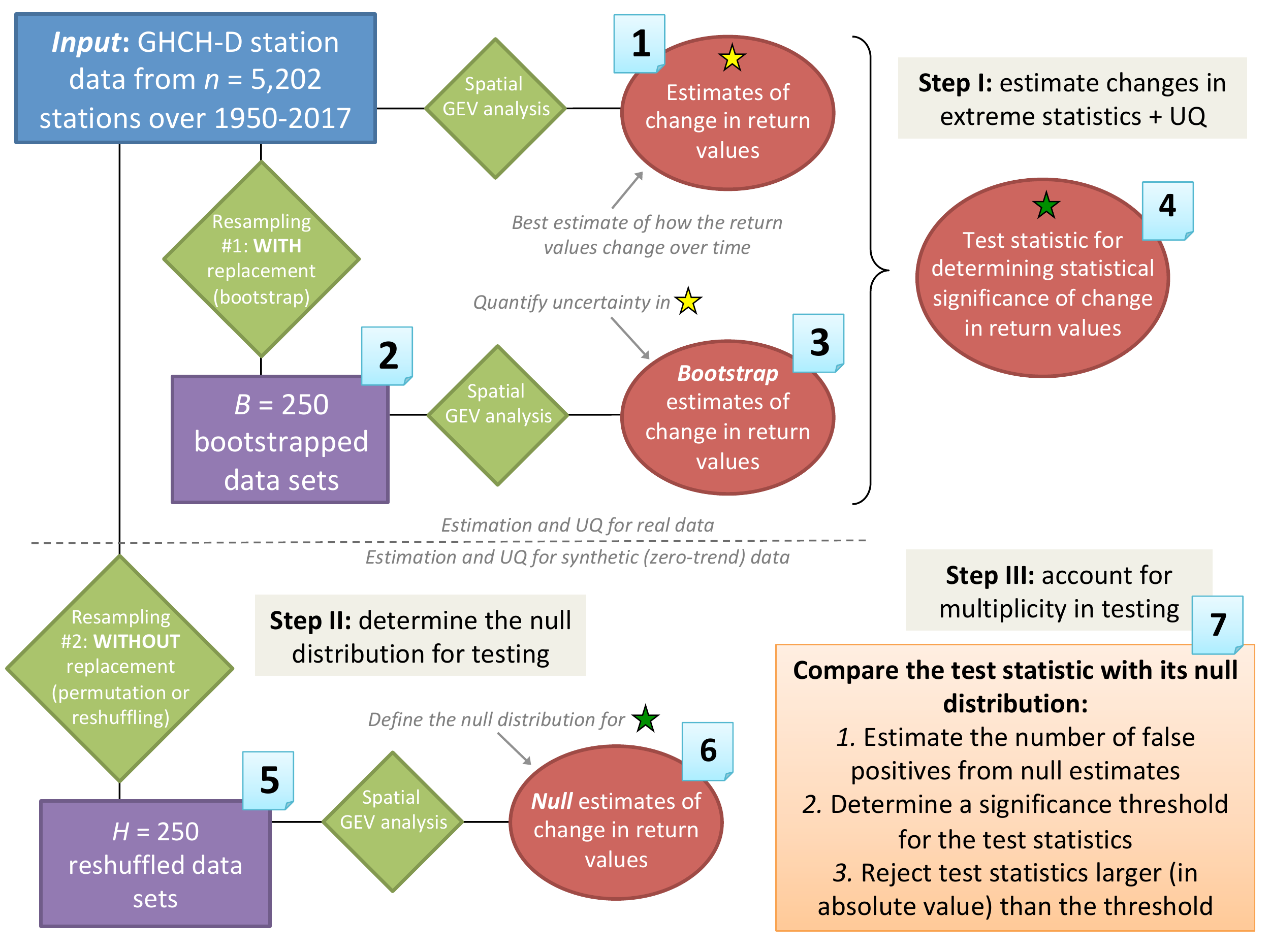}
\caption{A workflow diagram for the analysis described in Section \ref{section32}. In the diagram, dark blue boxes represent raw input data; purple boxes represent derived data sets; green diamonds represent operations (e.g., ``Spatial GEV analysis" refers to Section \ref{section31}); red circles represent an output. The numbered light blue boxes are pointers to specific components of the analysis and are referenced throughout Section \ref{section32}. In the gray boxes, Steps I-III correspond to the subsections of Section \ref{section32}.}
\label{workflow}
\end{center}
\end{figure}

The various components of the statistical methods used in our analysis are now described. Furthermore, the corresponding workflow is displayed schematically in Figure \ref{workflow} with specific pointers for each of the following steps.

\subsection{Spatial extreme value analysis} \label{section31}

In order to characterize the spatially-complete climatological distribution of extreme precipitation based on measurements from irregularly observed weather stations, we apply the spatial data analysis outlined in \cite{risser2018probabilistic}. In the workflow shown in Figure \ref{workflow}, this analysis is represented by the green diamond labelled ``Spatial GEV analysis.'' An important feature of the analysis is that it allows one to estimate the distribution of extreme precipitation even for locations where no data are available. Furthermore, the methodology proposed in \cite{risser2018probabilistic} can be used for a large network of weather stations over a heterogeneous spatial domain like CONUS, which is critical for the problem at hand. 

For a full description of the methodology used, we refer the reader to \cite{risser2018probabilistic}. In short, the essence of the method is to first obtain estimates of the climatological features of extreme precipitation based on measurements from the weather stations via the Generalized Extreme Value (GEV) family of distributions, which is a modeling framework for the maxima of a process over a pre-specified {time interval or} ``block,'' {e.g., the} three-month season{s used here}. 
\cite{Coles2001} (Theorem~3.1.1, page~48) shows that when the number of daily measurements is large (e.g., when considering {the} $\approx 90$ daily {measurements} {in a given} season), the cumulative distribution function (CDF) of $Y_t(\bfs)$ (which is the seasonal maximum daily precipitation measurement in year $t$ at station~$\bfs$) is a member of the GEV family
\begin{equation} \label{gev_fam}
G_{\bfs, t}(y) \equiv \mathbb{P}(Y_t(\bfs) \leq y) = \exp\left\{-\left[ 1 + \xi_t(\bfs)\left(\frac{y - \mu_t(\bfs)}{\sigma_t(\bfs)}\right) \right]^{-1/\xi_t(\bfs)} \right\}, 
\end{equation}
defined for $\{ y: 1 + \xi_t(\bfs)(y - \mu_t(\bfs))/\sigma_t(\bfs) > 0 \}$. The GEV family of distributions~(\ref{gev_fam}) is characterized by three space-time parameters: the location parameter $\mu_t(\bfs) \in \mathcal{R}${, }which describes the center of the distribution{;} the scale parameter $\sigma_t(\bfs)>0$, which describes the spread of the distribution{;} and the shape parameter $\xi_t(\bfs) \in \mathcal{R}$. The shape parameter $\xi_t(\bfs)$ is the most important {for} determining the qualitative behavior of the distribution of daily rainfall at a given location{. I}f $\xi_t(\bfs)<0$, the distribution has a finite upper bound; if $\xi_t(\bfs) >0$, the distribution has no upper limit; {and} if $\xi_t(\bfs) = 0$, the distribution is again unbounded and the CDF~(\ref{gev_fam}) is interpreted as the limit $\xi_t(\bfs) \rightarrow 0$ \citep{Coles2001}. 

As described in Section \ref{section1}, recall that the goal of this analysis is to characterize how the climatological distribution of extremes changes from 1950 to 2017. As such, we use the following time trend models for the GEV parameters:
\begin{equation} \label{coef_model}
\mu_t(\bfs) = \mu_0(\bfs) + \mu_1(\bfs) t, \hskip2ex \sigma_t(\bfs) \equiv \sigma(\bfs), \hskip2ex \xi_t(\bfs) \equiv \xi(\bfs)
\end{equation}
\citep[following, e.g.,][and others]{Westra2013,risser2018probabilistic}. We henceforth refer to $\mu_0(\bfs)$, $\mu_1(\bfs)$, $\sigma(\bfs)$, and $\xi(\bfs)$ as the \textit{climatological coefficients} for location $\bfs$, as these values describe the climatological distribution of extreme precipitation in each year. Of course, using only four parameters to model $G_{\bfs, t}(y)$ over time for each location is an admittedly simplistic representation given that the true smoothed trend could be non-linear (due to, e.g., the nonlinear trends in greenhouse gas concentrations). However, we argue that a linear approximation is reasonable, particularly since we evaluate estimated differences in $G_{\bfs, t}(y)$ between the endpoints of the time series (1950 and 2017; see Section \ref{section32}). We explored the possibility of (a) including nonlinear trends in the location parameter and (b) further modeling the scale and/or shape parameter as changing linearly over time; however, we found that the model described by (\ref{coef_model}) performed as well (in a statistical sense, i.e., using the Akaike information criteria; see Tables B3 and B4 in Appendix B for these comparisons) than any of these alternative characterizations. We further discuss nonlinear (quadratic) trends in Section \ref{sec:discussion}.
For all of these reasons, we argue that Eq.~(\ref{coef_model}) is a simplistic but acceptable model for characterizing smooth changes over 1950-2017 in the climatological distribution of extremes over CONUS.

Conditional on station-specific estimates of of the climatological coefficients, we next apply a spatial statistical approach using second-order nonstationary Gaussian processes to infer the underlying climatology over a fine grid via kriging. This approach yields fields of best estimates of the climatological coefficients, denoted $\{ \widehat{\mu}_0(\bfs'), \widehat{\mu}_1(\bfs'), \widehat{\sigma}(\bfs'), \widehat{\xi}(\bfs') : \bfs' \in \mathcal{G} \}$, where $\mathcal{G}$ is the $0.25^\circ$ grid of $M=13073$ grid cells over CONUS. These best estimates can be used to calculate corresponding estimates of the $r$-year return value, denoted $\widehat{\phi}^{(r)}_t(\bfs')$, which is defined as the seasonal maximum daily precipitation total that is expected to be exceeded on average once every $r$ years. In other words, $\widehat{\phi}^{(r)}_t(\bfs')$ is an estimate of the $1-1/r$ quantile of the distribution of seasonal maximum daily precipitation in year $t$ at grid cell $\bfs'$, i.e.,
$P\big(Y_{t}(\bfs') > \widehat{\phi}^{(r)}_t(\bfs')\big) = {1}/{r}$, which can be written in closed form in terms of the climatological coefficients:
\begin{equation} \label{returnVal}
\widehat{\phi}^{(r)}_t(\bfs') = \left\{ \begin{array}{ll}
[\widehat{\mu}_0(\bfs') + \widehat{\mu}_1(\bfs')t] - \frac{\widehat{\sigma}(\bfs')}{\widehat{\xi}(\bfs')}\big[1 - \{-\log(1-1/r)\}^{-\widehat{\xi}(\bfs')}\big],  & \widehat{\xi}(\bfs) \neq 0 \\[1ex]
[\widehat{\mu}_0(\bfs') + \widehat{\mu}_1(\bfs')t] - \widehat{\sigma}(\bfs') \log\{-\log(1-1/r)\},  & \widehat{\xi}(\bfs') = 0
\end{array} \right. 
\end{equation}
\citep{Coles2001}. Note that $\widehat{\phi}^{(r)}_t(\bfs')$, the best estimate of the return value in year $t$, does not represent a return value estimated from a single year of data but is instead calculated using climatological coefficients estimated from the complete time record with the specific year ($t$ in Eq.~\ref{returnVal}) plugged in. In other words, the time-varying return value estimates represent temporally smoothed quantities.

\subsection{Quantifying changes in extreme precipitation and significance testing} \label{section32}

The spatial extreme value analysis described in Section \ref{section31} is used to both quantify changes in the climatology of extreme precipitation and conduct significance testing to determine where meaningful changes in the distribution of extreme precipitation have occurred. The workflow described in the next three subsections is displayed schematically in Figure \ref{workflow}.

\subsubsection{Change metrics, test statistics, and null hypotheses}

First, we determine the metrics for evaluating changes in the return values, the relevant test statistics, and the null hypotheses of interest.
Using the results of the fitted and smoothed GEV distributions, we are interested in identifying areas that have experienced significant changes in the extreme statistics of precipitation{, }e.g., {the} return values{,} over 1950-2017. To quantify these changes, we consider the relative change in 20-year return values,  
\[
\Delta^R(\bfs') \equiv \frac{\phi^{(20)}_{2017}(\bfs') - \phi^{(20)}_{1950}(\bfs')}{\phi^{(20)}_{1950}(\bfs')},
\]
where $\phi^{(20)}_{t}(\bfs')$ is the $20$-year return value at grid cell $\bfs' \in \mathcal{G}$ in year $t$. Alternatively, we also consider the absolute change in 20-year return values
\[
\Delta^A(\bfs') \equiv {\phi^{(20)}_{2017}(\bfs') - \phi^{(20)}_{1950}(\bfs')}.
\]
Recall that the return values in $\Delta^R(\bfs')$ and $\Delta^A(\bfs')$ refer to temporally smoothed quantities evaluated in 1950 and 2017, not return value estimates specific to each year. 
Using the linear trend model in (\ref{coef_model}), note that $\Delta^R(\bfs')$ depends on the return period $r$, which is a necessary implication of only allowing the center of the distribution to shift over time: the relative change depends on where you are in the distribution. On the other hand, $\Delta^A(\bfs')$ does \textit{not} depend on $r$: given a constant shift per time in the center of the distribution only, all quantiles change by the same magnitude (in mm). Note that under the alternate parameterization of \cite{wiel2017rapid} (or any other parameterization that allows the scale parameter $\sigma_t$ to change over time, e.g., using a trend or physical covariates), both the relative and absolute changes depend on the return period $r$ (see Appendix A for further details). 

Let $\Delta(\bfs')$ represent an arbitrary change metric, which we assume to have some fixed but unknown value. Note the distinction between the true but unknown quantity $\Delta(\bfs')$, a function of the unknown $\phi^{(20)}_{t}(\bfs')$, and our estimate of this quantity $\widehat{\Delta}(\bfs')$ based on estimated return values $\widehat{\phi}^{(20)}_t(\bfs')$. (These estimates are represented in the flow diagram in Figure \ref{workflow} with pointer 1.) To test for the presence of any change in $\Delta(\bfs')$, {the null and alternative hypothesis for each location is}
\[
H_0(\bfs'): \Delta(\bfs') = 0 \hskip3ex \text{vs.} \hskip3ex H_1(\bfs'): \Delta(\bfs') \neq 0.
\]

The test statistic we use to assess the statistical significance of each $H_0(\bfs')$ is
\begin{equation} \label{Zscore}
z(\bfs') = \frac{\widehat{\Delta}(\bfs')}{\text{se}[\widehat{\Delta}(\bfs')]},
\end{equation}
where $\text{se}[\widehat{\Delta}(\bfs')]$ is the standard error of our best estimate $\widehat{\Delta}(\bfs')$. To estimate this uncertainty, we use 
%
the block bootstrap \citep[following, e.g.,][]{risser2018probabilistic} which,
as described in  \cite{risser2018probabilistic}, 
is important for quantifying uncertainty here because the two-stage spatial GEV analysis in Section \ref{section31} does not explicitly account for the spatial dependence in the daily measurements of precipitation, or the so-called ``storm dependence'' \citep[dependence due to the spatial coherence of storm systems; see][for further details]{risser2018probabilistic}. Instead of explicitly accounting for this systematic source of error, we rely on the block bootstrap 
to implicitly characterize the resulting uncertainty, since any real spatial signal will show up in most of the resampled data sets. Exploratory analyses in \cite{risser2018probabilistic} verify that this approach appropriately accounts for storm dependence.

The standard error $\text{se}[\widehat{\Delta}(\bfs')]$ is obtained as follows: recall ${\bf Y}(\bfs_i) = \{ Y_{t}(\bfs_i) : t = 1950, \dots, 2017 \}$ is the observed vector of the seasonal maxima weather station $\bfs_i$ (for now we suppress the $j$ subscript denoting the specific season). Then, for $b=1, \dots, B=250$, the bootstrap sample is obtained by drawing $T = 68$ years \textit{with replacement} from $\{ 1950, \dots, 2017 \}$, denoted $\{a_1, a_2, \dots,a_{T} \}$, so that the $b$th bootstrap sample for station $\bfs$ is ${\bf Y}_{b}(\bfs_i) = \big( Y_{a_1}(\bfs_i), \dots, Y_{a_T}(\bfs_i)  \big)$ (note that the temporal covariate values are resampled correspondingly). Furthermore, note that for a specific bootstrap sample,
all stations use the same sequence of years in order to maintain the spatial coherence of the underlying seasonal maxima. (Bootstrap sampling with replacement is represented in the flow diagram in Figure \ref{workflow} with pointer 2.) For each bootstrap sample, we then use the spatial extreme value analysis outlined in Section \ref{section31} to obtain the bootstrap estimates of the return values $\widehat{\phi}^{(r)}_{b,t}(\bfs')$ and change metric $\widehat{\Delta}_{b}(\bfs')$ for all $\bfs' \in \mathcal{G}$. (This step is represented in the flow diagram in Figure \ref{workflow} with pointer 3.) The resulting standard error is then
\[
\text{se}[\widehat{\Delta}(\bfs')] = \sqrt{ \frac{1}{B-1} \sum_{b = 1}^B \left(\widehat{\Delta}_{b}(\bfs') - \frac{1}{B}\sum_{b = 1}^B \widehat{\Delta}_{b}(\bfs')
\right)^2}.
\]
(This standard error is used to calculate the test statistic in Eq.~\ref{Zscore}; shown in the flow diagram in Figure \ref{workflow} with pointer 4.)

\subsubsection{Determination of null distribution for testing}

Next, we determine the null distributions against which to test the null hypotheses defined in the previous section.
In general, there are a variety of ways to use resampling-based procedures{, for example} the bootstrap{,} and the resulting $z$-scores (\ref{Zscore}) to conduct significance testing \citep[see, e.g.,][]{efron1994introduction}. However, in this case the test statistic (\ref{Zscore}) must be compared with corresponding $z$-scores under the assumption that the null hypothesis $H_0(\bfs')$ is true: in other words, we require a null distribution for the $z$-scores based on data that are known to have a trend of zero. Therefore, we cannot use the bootstrap estimates, because these are based on the original weather station data, which may or may not have a temporal trend. To navigate this issue, we can instead use a nonparametric permutation framework, in which we reshuffle the years in order to build up a data-driven null distribution for the $z$-scores based on data sets that have no trend by construction \citep[see Chapter 15 of][]{efron1994introduction}--as is also done in, e.g., \cite{lehmann2015increased}. This null distribution explicitly accounts for uncertainty arising from multiple sources: a limited sampling period, noisy measurement of the seasonal maxima, and the estimation procedure outlined in Section \ref{section31}.  The permutation test proceeds similarly to the bootstrap: 
\begin{enumerate}
\item For replicate $h = 1, \dots, H = 250$, 
\begin{enumerate}
 \item \textit{Without} replacement, sample ${d}_t \in \{ 1, \dots, T \}$, for $t = 1, \dots, T$. For each $\bfs_i \in \mathcal{S}$, collect the corresponding seasonal maxima into a vector, ${\bf Y}_{h}(\bfs_i) = \big( Y_{d_1}(\bfs_i), \dots, Y_{d_T}(\bfs_i) \big)$. As with the bootstrap, all stations use the same sequence of years to preserve the spatial structure of each reshuffling. (This step is represented in the flow diagram in Figure \ref{workflow} with pointer 5.)

\item Using the analysis outlined in Section \ref{section31}, obtain the best estimate of the field of test statistics, denoted $\{ \widehat{\Delta}_h(\bfs'): \bfs' \in \mathcal{G} \}$, based on the permuted data set $\{{\bf Y}_{h}(\bfs_i): \bfs_i \in \mathcal{S} \}$. Note that we treat these seasonal maxima for each weather station as if they arose from a consecutive $T$-year period. In other words, the covariate values are not shuffled in the temporal trend, unlike the bootstrap. (This step is represented in the flow diagram in Figure \ref{workflow} with pointer 6.)
\end{enumerate}
To be clear, note that these permutation estimates $\{ \widehat{\Delta}_h(\bfs'): h = 1, \dots, H \}$ define the null distribution for our best estimate of the change in the original (unshuffled) data at $\bfs'$, denoted $\widehat{\Delta}(\bfs')$. 

\item The permutation $z$-score for location $\bfs'$ is calculated as
\[
z_h(\bfs') = \frac{\widehat{\Delta}_h(\bfs')}{\text{se}[\widehat{\Delta}_h(\bfs')]}.
\]
In principle, one could bootstrap each reshuffled data set to estimate the standard error $\text{se}[\widehat{\Delta}_h(\bfs')]$. Instead, to avoid an unnecessary computational burden, we approximate the standard error for permutation sample $h$ using the standard deviation of the estimates from all $H=250$ permuted data sets
\begin{equation} \label{permSE}
\text{se}[\widehat{\Delta}_h(\bfs')] \approx \sqrt{ \frac{1}{H-1} \sum_{j= 1}^H \left(\widehat{\Delta}_{j}(\bfs') - \frac{1}{H}\sum_{j = 1}^H \widehat{\Delta}_{j}(\bfs')
\right)^2},
\end{equation}
which we verified gives quantitatively similar results to the brute-force approach of bootstrapping each permutation data set. 

\end{enumerate}

\subsubsection{Addressing multiplicity and field significance}

Finally, we deal with issues of multiplicity in hypothesis testing and field significance. (The steps that follow are represented in the flow diagram in Figure \ref{workflow} with pointer 7.)
Given that we are conducting a hypothesis test for each of $M = 13073$ grid cells in $\mathcal{G}$ (for each season), testing each hypothesis individually at a significance level of $\alpha$ (e.g., often $\alpha = 0.05$ or $0.10$) means that we would incorrectly reject $M\alpha$ of these hypotheses on average even if all $M$ null hypotheses were true. We must therefore think carefully about a multiple testing adjustment in order to avoid making a large number of Type I errors (i.e., false positives or false discoveries). Following arguments described in \cite{Risser2018a}, we seek to control the ``false discovery rate'' (FDR), or the proportion of Type I errors from the set of spatial locations for which we conclude there is a nonzero change. 

Usually, we might argue from a Bayesian or shrinkage perspective that the spatial smoothing underlying the calculation of the $z$-values would automatically account for the multiple testing aspect \citep{Gelman2008}, and we could simply compare $z(\bfs')$ with the corresponding $\{z_h(\bfs')\}$ to determine our decision regarding $H_0(\bfs')$. However in this case, our best estimates of the return values (and therefore change estimates) include both the true spatial signal as well as spatially-correlated error due the presence of storm dependence \citep[see][]{risser2018probabilistic}, and so we require a multiple testing adjustment even after borrowing strength over space. Any such adjustment must account for the fact that there is strong spatial correlation among the $M$ hypotheses. While most traditional multiple testing approaches are defined for independent hypotheses \citep[see, e.g.,][]{BenjaminiHochberg1995}, the literature also contain a number of approaches specifically designed for correlated hypotheses \citep[see, e.g.,][]{BenjaminiYekutieli2001,Pacifico2004,SunCai2009,Schwartzman2011,Sun2015,Shu2015,Risser2018a}. As discussed in \cite{Risser2018a}, many of these approaches are designed for specific statistical modeling frameworks and are not robust to deviations from the statistical model; on the other hand, we cannot use the robust Bayesian methodology of \cite{Risser2018a} since we are in a Frequentist setting. 

While \cite{BenjaminiHochberg1995} was originally derived for independent hypotheses, their procedure is also valid (in the sense of controlling the FDR) under positive regression dependency (PRD) of the test statistics \citep{BenjaminiYekutieli2001}. Intuitively, PRD is satisfied if the test statistics are positively correlated \citep[see Section 2.2 of][]{BenjaminiYekutieli2001}; one can argue that this condition holds here when dealing with spatially smooth change estimates. Therefore, we can use an argument similar to the one outlined in Section 3.3.3 of \cite{BenjaminiHochberg1995} to derive our testing procedure: here, the authors frame their FDR-controlling procedure as a \textit{post hoc} maximization where one finds the value $\alpha \in [0, 1]$ that maximizes the number of rejections $R(\alpha)$ subject to the constraint that 
\begin{equation} \label{constraint}
\alpha M/R(\alpha) \leq q,
\end{equation}
where $q$ is the desired rate of false discoveries. Note that this procedure is implicitly based on a set of $P$-values, e.g., $\{ p(\bfs'): \bfs' \in \mathcal{G}\}$, in that $R(\alpha) = \sum_{\bfs' \in \mathcal{G}} I\{ p(\bfs') \leq \alpha \}$ (here, $I\{\cdot\}$ is an indicator function) as described in \cite{BenjaminiHochberg1995}. Therefore, after observing the outcome of the procedure, the value $\alpha$ can be chosen by maximizing the number of rejections subject to the constraint in (\ref{constraint}). Framing the FDR procedure in this way is based on the fact that $V(\alpha)$, the expected number of false rejections for a particular $\alpha$, can be bounded above by $\alpha M$ because $V(\alpha) = \alpha M$ when all of the null hypotheses are true. 

However, given the permutation results, we can adapt this framework slightly to yield a more appropriate FDR procedure. First, unlike \cite{BenjaminiHochberg1995}, we define a rejection cutoff in terms of $z$-scores (instead of $P$-values), because we do not have a defensible method for deriving appropriate $P$-values from the permutation estimates. As such, the number of rejections and false rejections are now defined in terms of a $z$-score cutoff $c$, i.e., $R(c)$ and $V(c)$. Second, we do not need to assume a theoretical result for the number of false rejections (in the previous paragraph, this was $V(\alpha) = \alpha M$) because we can derive an estimate empirically from the permutation $z$-scores $\{ z_h(\bfs'): h = 1, \dots, H \}$. These $z$-scores automatically account for both the spatial correlation among the tests and all sources of uncertainty in the analysis. For each $h$, the number of false rejections for a given $z$-score cutoff $c \geq 0$ is estimated by 
\begin{equation} \label{V_alpha}
\widehat{V}_h(c) = \sum_{\bfs' \in \mathcal{G}} I\big\{ |z_h(\bfs')| > c \big\}
\end{equation}
because the $z$-scores $\{z_h(\bfs')\}$ are calculated from a reshuffled data set with no change and hence none of the tests should be rejected. Then, averaging over all permutation data sets, an estimate of the overall number of false rejections for a given $c$ is $\widehat{V}(c) = H^{-1} \sum_{h =1}^H \widehat{V}_h(c)$, which is an improvement on the theoretical upper bound used in \cite{BenjaminiHochberg1995} in that it accounts for the spatial correlation of the collective set of hypotheses. Our testing procedure that includes a multiple testing adjustment proceeds as follows (these three steps are represented in the flow diagram in Figure \ref{workflow} with pointer 7):
\begin{enumerate}
\item For each $c \geq 0$, calculate $\widehat{V}(c)$ following (\ref{V_alpha}) as well as the number of rejections in the full data, $\widehat{R}(c) = \sum_{\bfs' \in \mathcal{G}} I\{ |z(\bfs')| > c \}$.

\item Find the smallest value $c$ such that $\widehat{V}(c)/\widehat{R}(c) \leq q$, denoted $c^*$.

\item Reject all $H_0(\bfs')$ with $|z(\bfs')| > c^*$.
\end{enumerate}

Applying this procedure ensures that the rate of false discoveries (equivalently, the rate of type I errors or false positives) is bounded by $q$ \citep{BenjaminiHochberg1995}. This yields a systematically different characterization of statistical significance than the concept of field significance \citep[FS;][]{LivezeyChen1983}, which is an alternative multiple testing approach that seeks to evaluate the collective significance of a set of statistics. As such, FS provides no specific information about which individual grid cells are significant. A procedure that controls the FDR, on the other hand, does provide this local information: we can be confident that any grid cell flagged as significant is indeed significant. In any case, while we argue that FDR-control provides a better characterization of statistical significance, we also present the FS of the changes for comparison, as this provides an alternate and complementary line of evidence for the collective significance of the changes. Here, the field significance is calculated based on the original data $z$-scores $\{z(\bfs')\}$ and the permutation $z$-scores $\{z_h(\bfs')\}$ as follows: first, calculate the cumulative distribution functions (CDFs) of the absolute value of the $z$-scores, denoted $F(z)$ and $F_h(z)$. Then, for each $z$-score cutoff $c$, the field significance is calculated as
\begin{equation} \label{fs}
FS(c) = \frac{1}{H} \sum_{h=1}^H I\{ F_h(c) < F(c)\};
\end{equation}
i.e., the proportion of the permutation CDFs that are less extreme than the original data CDF (in other words, the permutation probability density function has a heavier upper tail relative to the full data density).  

\section{Results} \label{sec:results}

\subsection{Seasonal results}

\begin{figure}[!t]
\begin{center}
\includegraphics[trim={0 0 0 0mm}, clip, width = \textwidth]{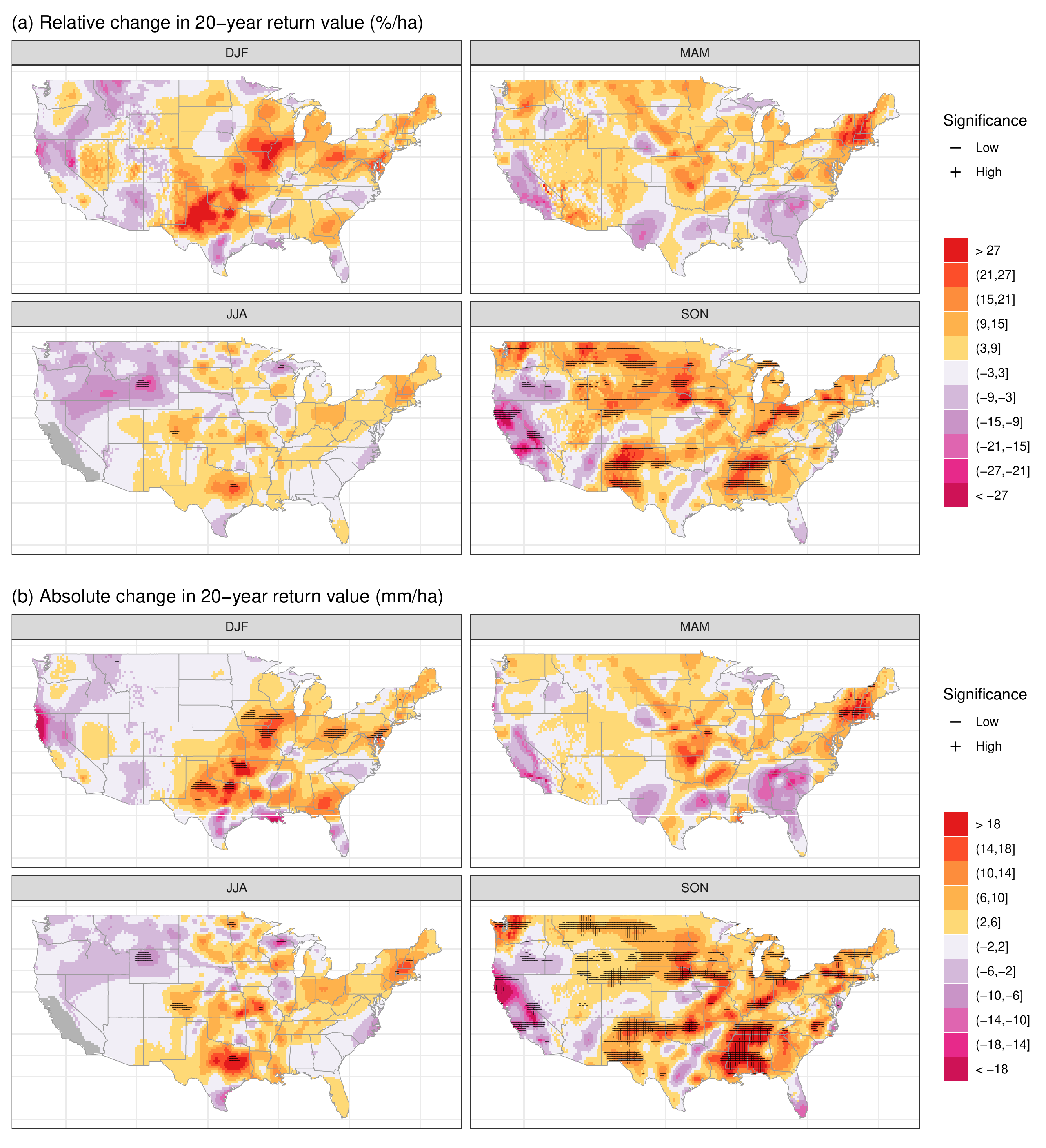}
\caption{Gridded best estimates of the (a) relative change (percent per century or ha) and (b) absolute change (mm per ha) in the 20-year return value (1950 vs. 2017). Grid cells hashed with a ``$-$'' contain a change with low statistical significance (i.e., the proportion of type I errors for these grid cells is $\leq 33\%$); those with a ``$+$'' have high statistical significance (i.e., the proportion of type I errors for these grid cells is $\leq 10\%$).}
\label{figure2}
\end{center}
\end{figure}

Applying the analysis procedure described in Section \ref{StatMethods} to the seasonal maxima, our best estimates of the relative change in the $20$-year return value (i.e., $\Delta^R(\bfs')$) are shown in Figure \ref{figure2}(a) with best estimates of the absolute change (i.e., $\Delta^A(\bfs')$) in Figure \ref{figure2}(b). Note that part of California is masked out for JJA because the dry season in this region means that an extreme value distribution for the seasonal maxima at these stations is inappropriate. 
In general, the largest spatially-coherent changes appear in DJF, SON, and possibly MAM. The dominant direction of the observed changes is positive, although there are also large regions with negative changes in each season. 

Figure \ref{figure2} also displays the statistical pointwise significance of these changes corresponding to ``low'' (i.e., controlling the rate of false discoveries at $q = 0.33$) and ``high'' (i.e., controlling the rate of false discoveries at $q = 0.1$) confidence statements. 
These thresholds for significance are chosen as reasonable limits for bounding the proportion of type I errors, with $q = 0.33$ yielding a less conservative statement and $q = 0.1$ yielding a more conservative statement.
The proportion of the map falling into each significance category for each season is given in Table \ref{TAB_SUMMARY} (in the ``Gridded'' columns). With regards to the relative change in return value, very few of the grid cell changes are significant even at the low level for DJF, MAM, and JJA; in SON, on the other hand, a large proportion of CONUS (21.9\%) contains changes with a low level of significance, with approximately 19.5\% of these being increasing changes and the remaining 2.4\% being decreasing changes. Turning to the absolute changes in return value, all seasons now exhibit at least some areas with changes that are at least of low significance, with spatially-coherent increasing changes in the central United States in DJF and the New England states in MAM. As with the relative change in return value, a large proportion of CONUS display changes with a low significance level in SON (33.3\% of the grid cells), with 6.8\% of the grid cells further showing highly significant changes. Of those with low significance in SON, there are spatially coherent increases in the southeast US, New Mexico, and Northern Great Plains; on the other hand, there are highly significant decreases in California (specifically the northern coast and a cluster in the Sierra Nevada).

\begin{table}[!t]
\caption{Proportion of independent station estimates (``Independent'') and spatially-smoothed grid cells (``Gridded'') falling into each confidence category for (a) relative change in return value and (b) absolute change in return value in each season.}
\begin{center}
\begin{tabular}{|p{0.15\linewidth}|| p{0.15\linewidth}|  p{0.15\linewidth} || p{0.15\linewidth} | p{0.15\linewidth}|} \hline

\multicolumn{5}{|l|}{\textbf{(a) Relative change in return value} } \\ \hline\hline
 & \multicolumn{2}{|c||}{\textit{Low significance (only)}} & \multicolumn{2}{|c|}{\textit{High significance (only)}} \\ \hline
	 & Independent & Gridded & Independent & Gridded \\ \hline
DJF  & 0.0058 & 0.0000 & 0.0000 & 0.0000 \\ \hline
MAM  & 0.0004 & 0.0000 & 0.0000 & 0.0000 \\ \hline
JJA	 & 0.0010 & 0.0058 & 0.0000 & 0.0000 \\ \hline
SON  & 0.0031 & 0.2185 & 0.0000 & 0.0000 \\ \hline 


\multicolumn{5}{l}{} \\
\hline
\multicolumn{5}{|l|}{\textbf{(b) Absolute change in return value} } \\ \hline\hline
 & \multicolumn{2}{|c||}{\textit{Low significance (only)}} & \multicolumn{2}{|c|}{\textit{High significance (only)}} \\ \hline
	 & Independent & Gridded & Independent & Gridded \\ \hline
DJF  & 0.0686 & 0.0400 & 0.0258 & 0.0000 \\ \hline
MAM  & 0.0336 & 0.0121 & 0.0083 & 0.0000 \\ \hline
JJA	 & 0.0125 & 0.0169 & 0.0038 & 0.0000 \\ \hline
SON  & 0.0677 & 0.2655 & 0.0136 & 0.0675 \\ \hline


\end{tabular}
\label{TAB_SUMMARY}

\end{center}
\end{table}

Clearly, when considering both the relative and absolute change in return values, there are many grid cells with a large change (in absolute value) that are not statistically significant even at the lower level. In principle, this could be due to several factors: noise in the station data itself, an overly conservative false discovery rate procedure, or the statistical interpolation procedure used throughout. The lack of significance is not due to the choice of return level $r = 20$: the significance maps are approximately the same when considering other return levels, both more and less extreme (not shown). 
In this analysis, however, the predominant causes are (1) noise in the station data and (2) unaccounted-for storm dependence in the underlying precipitation measurements. First, consider Figure B1 (in the appendix) as well as Table \ref{TAB_SUMMARY} (in the ``Independent'' columns), which show the results of a corresponding analysis of changes in precipitation extremes at each station that does not borrow strength spatially (but still applies the multiple testing adjustment). Very few stations exhibit even low significance in the relative change in return value, and while there are a moderate number of stations with high significance in the absolute change in return value (approximately 3-6\% for each season), the changes do not exhibit the spatial coherence that might be expected, consistent with the analysis of \cite{Balling2011} indicating that station-based changes exhibit a high degree of entropy. This suggests that noise in the station data is the culprit, as opposed to the interpolation procedure itself. (Note: here, ``noise'' refers to both inherent uncertainty in the estimation of parameters in our GEV model as well as measurement errors that induce errors in the change estimates.)

A second contributing factor to the relatively few significant changes is the unaccounted-for dependence in the underlying fields of daily precipitation, or storm dependence. Returning to the gridded results shown in Figure \ref{figure2}, recall that the significance assessment is based on reshuffling or permuting the seasonal maxima. Even though these permuted data sets have no time trend by construction, it is possible that the estimated change is as large (or larger) than the change estimates shown in Figure \ref{figure2}. For example, Figure B2 in the appendix shows the standardized relative change in return value (as a $z$-score) in DJF for the unshuffled (original) data versus the corresponding quantity for five permutation data sets with particularly large changes. Clearly, the $z$-score for the change estimates in the permuted data sets can be of the same order as the $z$-score from the original data. As discussed in \cite{risser2018probabilistic}, the spatial extreme value analysis conducted here cannot distinguish between spatial correlation in the true underlying change and spatially-correlated error due to storm dependence. This is likely the reason why we estimate such large changes in the permutation data sets: the random reshuffling of individual years with an individual storm that impacted a large geographic area (and hence a large number of stations) could result in large (but spurious) positive or negative changes. 

To consider this more generally for all permutation data sets, consider Figures B3 and B4 in the appendix, which show the cumulative distribution functions (CDFs) for the relative and absolute change in return values (respectively). Specifically, these figures show the CDF of the absolute value of the $z$-scores for both the original (unshuffled) data as well as each of the individual permutation data sets. The unshuffled CDF is systematically larger than the permutation mean CDF across all $z$-score values for both the relative and absolute change, but (aside from SON) the unshuffled CDF is not extremely unusual relative to the individual permutation CDFs. Therefore, we have good reason to believe the false discovery rate procedure is not overly conservative because our testing results are in line with what we might intuitively expect from Figures B3 and B4.

Finally, the field significance of these changes is shown in Figure B5 (note: the calculations for $FS(c)$ in Equation \ref{fs} are based on the CDFs shown in Figures B3 and B4). For the relative change, SON has a strong signal emerging regardless of the significance threshold; otherwise, only JJA has a small degree of field significance. This matches the FDR results, which show a large number of significant grid cells in SON and only a few significant cells in JJA. For the absolute change, on the other hand, all seasons have strong signals emerging in terms of field significance, with SON showing the strongest signal. Again, this is in agreement with the FDR-based significance statements, for which all seasons have at least some grid cells with pointwise significance. We reiterate that the FS results here (and in general) do not inform \textit{where} the significant changes are, and furthermore FS tends to imply an over-confidence in the spatial extent of significant changes: as an example, for the absolute change, the JJA map is field significant even though only a small proportion of the grid cells (approximately 1\%) are pointwise statistically significant.

\subsection{Annual results}

While all results so far have been presented seasonally, we can also use the results from our fitted statistical model to make a statement about annual changes in extreme precipitation. Note that we do \textit{not} re-apply our spatial extreme value analysis to the annual maxima at each station: our analyses are conducted on the seasonal maxima because extreme precipitation results from different physical processes in different seasons, and aggregating over these differences confounds the important characteristics of the resulting extremes. Nonetheless, annual changes can be useful to water resource managers.

\begin{figure}[!t]
\begin{center}
\includegraphics[trim={0 0 0 0mm}, clip, width = 0.9\textwidth]{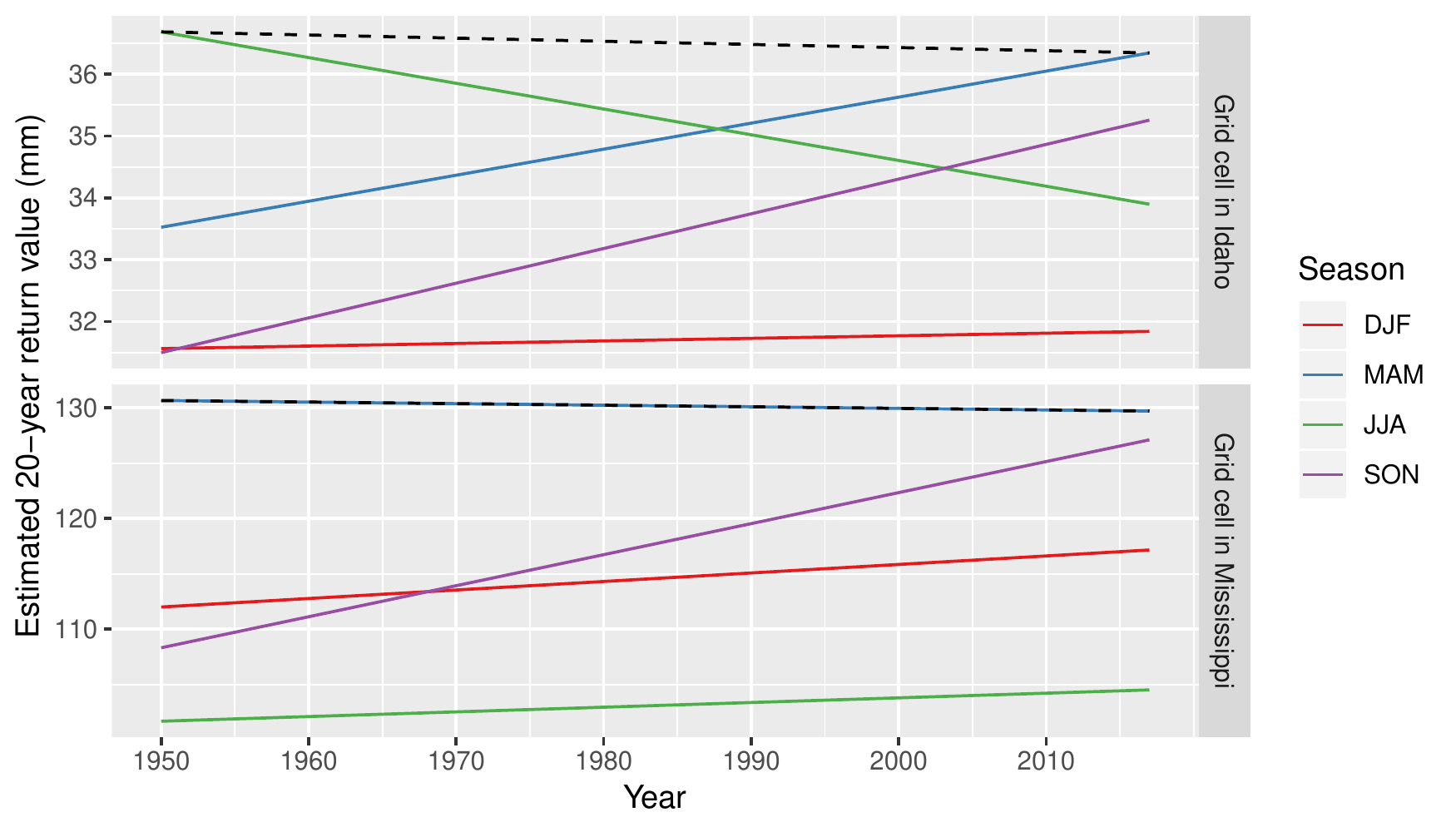}
\caption{Estimated 20-year return values (mm) for each season over 1950-2017, with the annual ``trend'' shown with the dashed black line. The top (bottom) panel corresponds to a selected grid cell in Idaho (Mississippi). Note that the annual trend for the grid cell in Mississippi coincides exactly with the MAM seasonal trend.}
\label{figure5}
\end{center}
\end{figure}

Here, we first conduct seasonal analyses, then calculate the largest 20-year return value amongst the four seasonal values in each year (which implicitly accounts for distributional differences) and then estimate a trend over time from the derived (artificial) time series. Given that we have specified a linear trend, this means that we can simply compare the largest seasonal return value in 2017 with the largest seasonal return value in 1950. In other words, we can assess
\begin{equation} \label{annualRC}
\widehat{\Delta}^{R}_*(\bfs') \equiv \frac{\max_u\{\widehat{\phi}^{(20)}_{u, 2017}(\bfs')\} - \max_u\{\widehat{\phi}^{(20)}_{u, 1950}(\bfs') \} }{\max_u\{\widehat{\phi}^{(20)}_{u, 1950}(\bfs')\}}
\end{equation}
and
\begin{equation} \label{annualAC}
\widehat{\Delta}^{A}_*(\bfs') \equiv \max_u\{\widehat{\phi}^{(20)}_{u, 2017}(\bfs')\} - \max_u\{\widehat{\phi}^{(20)}_{u, 1950}(\bfs')\}
\end{equation}
where $u = \{\text{DJF, MAM, JJA, SON}\}$ are the different seasons and $\widehat{\phi}^{(20)}_{u, t}(\bfs)$ is the \textit{estimated} 20-year return value in season $u$ and year $t$ at grid cell $\bfs$. As an illustration, consider Figure \ref{figure5} which shows the estimated seasonal trends for two sample grid cells as well as the annual ``trend''. Clearly, the annual change can be no larger than the largest seasonal change: note that for the grid cell in Idaho, the annual trend is nearly flat in spite of relatively large changes in MAM, JJA, and SON; on the other hand, the annual change can coincide with one of the seasonal changes, as in the grid cell in Mississippi.

\begin{figure}[!t]
\begin{center}
\includegraphics[trim={0 0 0 0mm}, clip, width = \textwidth]{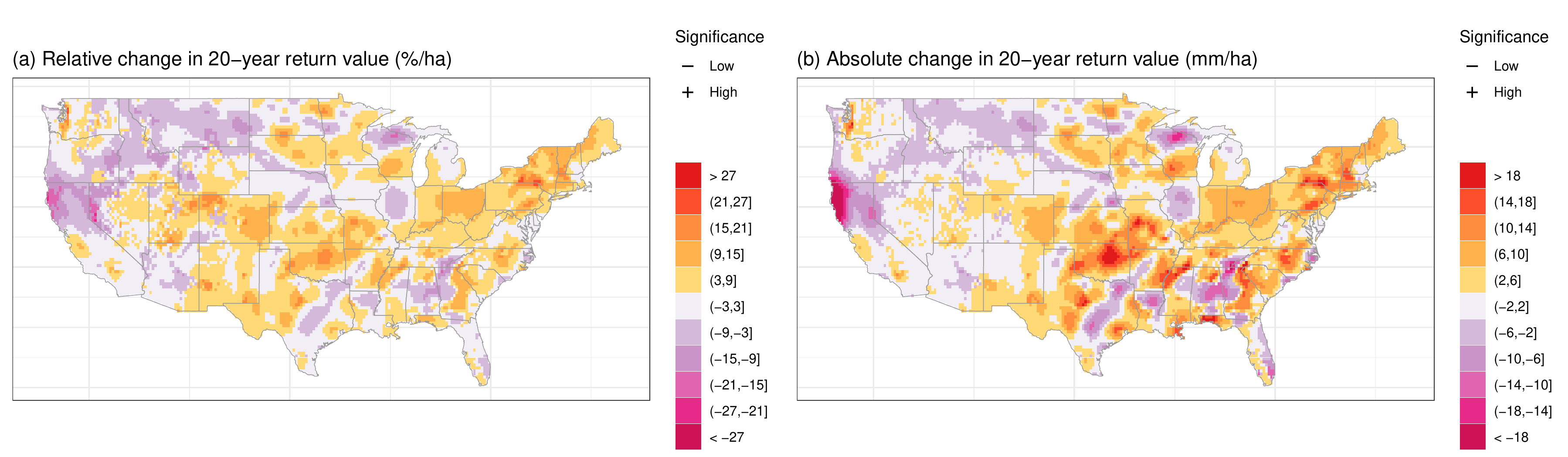}
\caption{Annual gridded best estimates of the (a) relative change and (b) absolute change in the 20-year return value (1950 vs. 2017). Grid cells hashed with a ``$-$'' contain a change with low statistical significance (i.e., the proportion of type I errors for these grid cells is $\leq 33\%$); those with a ``$+$'' have high statistical significance (i.e., the proportion of type I errors for these grid cells is $\leq 10\%$). Note that no cells in either panel are significant even at the lowest level.}
\label{figure4}
\end{center}
\end{figure}

Applying the same significance testing procedure as outlined in Section \ref{section32} to the annual changes $\widehat{\Delta}^{R}_*(\bfs')$ and $\widehat{\Delta}^{A}_*(\bfs')$, the best estimates of the annual relative and absolute changes are shown in Figure \ref{figure4}. Note that none of the annual changes are statistically significant even at the low significance level (nor are the maps field significant). This is not entirely surprising given the way these changes were constructed: again looking at Figure \ref{figure5}, note that a large SON change in Mississippi is disguised by a nearly flat trend in the wetter MAM season.

\section{Discussion} \label{sec:discussion}

In this paper, we have characterized changes in observed extreme precipitation for a fine spatial grid over CONUS using station data only and derived a statistical technique to robustly identify statistically significant pointwise changes in the climatological distribution of extreme precipitation. Our results constitute a novel contribution to the literature on observed changes in extreme precipitation because (1)~
we characterize changes at a point-based scale that is the native spatial scale for precipitation observations,
(2)~we account for seasonal differences, and (3)~we report uncertainty by ascribing statistical significance. No other study to our knowledge simultaneously addresses these three components (although, e.g., \citep{zhang2010influence} explore seasonal changes in extremes for individual stations). Furthermore, we describe an approach for constructing an annualized change based on the seasonal estimates that is of interest to, e.g., water resource managers.

The maps in Figure \ref{figure2}(b) exhibit a strong correspondence with the observed changes in 20-year return values over 1948 to 2015 given in \cite{Easterling2017} for seven large aggregated geographic regions \citep[see Figure 7.2 in][]{Easterling2017}. In winter, \cite{Easterling2017} find large positive changes in the southern Great Plains, with smaller increases in the Midwest, Southeast, and Northeast; our results identify the same changes. There is a similar agreement for the other seasons, e.g., large increases in the Northeast for spring, the southern Great Plains in summer, and the Southeast in fall. 
However, note that \cite{Easterling2017} show roughly zero change in the Southwest for fall, which is likely the result of aggregating over individual station trends (see Figure B1) that show large decreases in Northern California and correspondingly large increases in New Mexico.
These similarities provide support for the accuracy of our results while illustrating the impact (and importance) of resolving changes spatially.

The technique developed to produce this data set provides a natural method for evaluating changes in historical simulations at the native scale of the simulations.  The Gaussian process method used to model the spatial statistics of the station-based changes can be used to estimate the changes at arbitrary spatial locations: e.g., locations of model grid cells.  Model-based precipitation is typically a cell-averaged quantity, and so the magnitudes of extreme values will generally be lower than the point-based values used here.  Because of this, \citet{Chen2008} advocate for utilizing equivalently area-averaged observations when evaluating the statistics of simulated precipitation.  However, as model resolution increases, the values of precipitation should asymptote to point-like values, and so the changes from this data set (specifically, the significant changes) represent a quantity to which a suitable historical model should converge as resolution increases.  These two perspectives are not mutually exclusive: the \citet{Chen2008} perspective suggests that the most fair evaluation of model output would utilize precipitation data that is aggregated in a way that represents an appropriate area average, whereas our perspective suggests that point-based values provide statistics to which a model should converge as grid spacing approaches zero.  This effectively represents an upper bound for the statistics of extremes from simulations.   This data set could therefore form the basis for a complementary metric of model performance in simulating temporal changes in extremes, and this metric could be tracked over model development cycles.  More broadly, it is worth considering whether the methodology employed here could be merged with that of \cite{Donat2013} to produce spatially complete estimates of a variety of metrics of extremes: at the native resolution of models against which the data would be compared.

Of course, our analysis involves several important limitations. While we have carefully derived a valid statistical procedure for detecting statistically significant changes, note that our results appear to be conservative since we emphasize the pointwise significant results and not the global assessment of field significance. As mentioned in Section \ref{sec:results}, the changes might be clarified (or ``more significant'') if we accounted for the spatial dependence in the underlying fields of daily precipitation, but there is no existing methodology to do so appropriately for a large network of weather stations over a heterogeneous domain like CONUS. A potentially clearer way to handle the multiple testing would be to use the methodology in \cite{Risser2018a}--in a Bayesian setting, this would allow us to answer questions like ``is the relative change in return value greater than $5\%$ per century?''--but again the lack of appropriate methodology prevents us from using such an approach in this paper.

It should also be noted that the Generalized Extreme Value (GEV) distribution used throughout this paper is the asymptotic (limiting) distribution of the sample maximum of a large sample of $n$ independent draws from a parent population; this theoretical result applies exactly as the sample size $n\rightarrow\infty$ \citep[see, e.g.][Chapter 3]{Coles2001}. Of course, in practice, one only has a finite sample size; furthermore, in a block maxima framework, there is a trade-off between choosing large block sizes (leading to a better approximation, i.e., less bias, but larger variance due to fewer block maxima) versus small block sizes (an increase in the bias but smaller variance). A standard choice when dealing with climate data is to choose block sizes of one year; in this paper, we have chosen to apply the GEV distribution to seasonal maxima, where $n = 90$ (in which case we have $T = 68$ blocks or years of data). While a block size of 90 (in general) might be considered sufficient for the asymptotic results to hold, there are two important considerations when analyzing real measurements of daily precipitation:
\begin{enumerate}
    \item In some locations over CONUS and in some seasons, there are a large number of days with zero precipitation. Technically, the GEV approximation applies to draws from any underlying parent population (e.g., a distribution with a point mass at zero and a long right tail), but the point remains that if we are considering the ``precipitation process'' (i.e., days with nonzero rainfall), we may have fewer than 90 measurements of this process in a 90-day season.
    \item Furthermore, the GEV approximation applies to a \textit{statistically independent} sample of size $n$. While daily precipitation may not exhibit strong temporal autocorrelation, the effective sample size of independent measurements of daily precipitation from a 90-day season is certainly less than 90.
\end{enumerate}
In light of these considerations, Appendix C contains the results of a simple perfect data experiment (or simulation study) to evaluate the quality of the GEV approximation as a function of sample (block) size. In summary, there is indeed a non-negligible effect of block size on the quality of the GEV approximation and resulting bias/uncertainty quantification for estimating return values, but when estimating return values for return periods that are not too extreme (within the range of the data) this effect is minimal. Certainly for the 20-year return values estimated in this paper, we can be confident that our estimates contain minimal bias and the standard errors are well-calibrated. This result holds even when the effective number of measurements of daily precipitation we have from each season (due to either zero rainfall days, autocorrelation, or missing data) is as low as 25 or even 10. It should be noted that this result is absolutely not true when considering much longer-term (e.g., 500- or 1000-year) return values: in this case there is significant bias and the standard errors are much too large (i.e., conservative).

Returning to the potential inappropriateness of using a linear trend to characterize non-smooth and possibly nonlinear trends over the last 70 years, we reiterate that the simple trend model used here nonetheless yields important insights into changes in the climatology of extreme precipitation. As a sensitivity analysis, we analyzed the same data using a quadratic trend in the location parameter, i.e., $\mu_t = \mu_0 + \mu_1t + \mu_2t^2$ (although still using $\sigma_t \equiv \sigma$ and $\xi_t \equiv \xi$). The resulting relative and absolute changes using a quadratic trend correspond very closely to the changes using a linear trend (see Figure B6 in the Appendix), which provides support for the the appropriateness of characterizing changes using a linear trend. We reiterate that when attempting to explain (i.e., attribute) these changes it will be necessary to account for interannual and decadal (natural) variability in precipitation extremes, which we will explore in future work. 

\singlespacing
\bibliographystyle{apalike} \bibliography{detected_changes_ref}

\begin{appendix}
\numberwithin{figure}{section}
\numberwithin{table}{section}
 
\section{Changes in return values} \label{app:chgRV}

To simplify notation in the following, note that when the shape parameter is constant over time, i.e., $\xi_t(\bfs) \equiv \xi(\bfs)$, we can write the $r$-year return value in year $t$ as 
\[
\phi^{(r)}_t(\bfs) = \mu_t(\bfs) - \sigma_t(\bfs) f_{\xi(\bfs)}(r),
\]
where 
\[
f_{\xi(\bfs)}(r) = \left\{ \begin{array}{ll}
\frac{1}{\xi(\bfs)}\big[1 - \{-\log(1-1/r)\}^{-\xi(\bfs)}\big],  & \xi(\bfs) \neq 0 \\[1ex]
\log\{-\log(1-1/r)\},  & \xi(\bfs) = 0.
\end{array} \right.
\]
Recall from Section \ref{section31} that the time-varying model we use in this paper is
\[
\mu_t(\bfs) = \mu_0(\bfs) + \mu_1(\bfs) t, \hskip2ex \sigma_t(\bfs) \equiv \sigma(\bfs), \hskip2ex \xi_t(\bfs) \equiv \xi(\bfs).
\]
For this model, the relative change in return value is
\[
\Delta^R(\bfs) = \frac{\phi^{(r)}_{t_2}(\bfs) - \phi^{(r)}_{t_1}(\bfs)}{\phi^{(r)}_{t_1}(\bfs)} =  \frac{\mu_1(\bfs)(t_2 - t_1)}{\mu_0(\bfs) + \mu_1(\bfs)t_1 - \sigma(\bfs) f_{\xi(\bfs)}(r)}
\]
which depends on $r$; on the other hand, the absolute change in return value is
\[
\Delta^A(\bfs) = \phi^{(r)}_{t_2}(\bfs) - \phi^{(r)}_{t_1}(\bfs) =  \mu_1(\bfs)(t_2 - t_1),
\]
which is independent of $r$. An alternative four-parameter representation of the time-varying model is
\[
\mu_t(\bfs) = \mu_0 \exp\left\{ \frac{\alpha(\bfs) t}{\mu_0(\bfs)} \right\}, \hskip3ex \sigma_t(\bfs)  = \sigma_0(\bfs) \exp\left\{ \frac{\alpha(\bfs) t}{\mu_0(\bfs)} \right\},  \hskip3ex \xi_t(\bfs) \equiv \xi(\bfs)
\]
(\citealp{wiel2017rapid}), which allows both the location and scale to vary over time such that their ratio is a constant, i.e., $\mu_t(\bfs)/\sigma_t(\bfs) = \mu_0(\bfs)/\sigma_0(\bfs)$. Under this parameterization, the relative change in return value is 
\[
\frac{ \left[ \mu_0(\bfs) \exp\left\{ \frac{\alpha(\bfs) t_2}{\mu_0(\bfs)} \right\}  - \sigma_0(\bfs) \exp\left\{ \frac{\alpha(\bfs) t_2}{\mu_0(\bfs)} \right\}f_{\xi(\bfs)}(r) \right] -  \left[ \mu_0(\bfs) \exp\left\{ \frac{\alpha(\bfs) t_1}{\mu_0(\bfs)} \right\}  - \sigma_0(\bfs) \exp\left\{ \frac{\alpha(\bfs) t_1}{\mu_0(\bfs)} \right\}f_{\xi(\bfs)}(r) \right]}{ \mu_0(\bfs) \exp\left\{ \frac{\alpha(\bfs) t_1}{\mu_0(\bfs)} \right\}  - \sigma_0(\bfs) \exp\left\{ \frac{\alpha(\bfs) t_1}{\mu_0(\bfs)} \right\}f_{\xi(\bfs)}(r) }
\]
which again depends on $r$; note that the numerator (which is the absolute change) also depends on $r$.

\section{Supplementary figures and tables}

\begin{table}[h]
\caption{Fixed quantities used throughout the analysis.}
\begin{center}
\begin{tabular}{|p{2.45cm}|p{12cm}|}
\hline
\textbf{Quantity and fixed value} & \textbf{Definition} \\ \hline\hline
$T = 68$    & The number of years (1950-2017) used in the analysis, also the number of seasonal maxima used to estimate extreme value statistics over time for each station. \\ \hline
$n = 5202$ & The number of reliable GHCN stations and their geographic locations. \\ \hline
$M = 13073$ & The number of $0.25^\circ$ degree grid cells over CONUS. \\
\hline
\end{tabular}
\end{center}
\label{fixed_pars}
\end{table}%

\begin{table}[h]
\caption{Specified quantities used throughout the analysis.}
\begin{center}
\begin{tabular}{|p{2.45cm}|p{12cm}|}
\hline
\textbf{Quantity and fixed value} & \textbf{Definition} \\ \hline\hline
$B = 250$    & The number of bootstrap data sets used to quantify all sources of uncertainty in the original (unshuffled) data. \\ \hline
$H = 250$    & The number of reshuffled or permuted data sets used to approximate the null distribution and quantify all sources of uncertainty. \\ \hline
$q^* = 0.33$ and $0.1$ & The acceptable rates of false discoveries, which also define the statistical significance thresholds. These correspond, respectively, to low and high confidence statements. \\ \hline
\end{tabular}
\end{center}
\label{free_pars}
\end{table}%

\begin{table}[t]
\caption{Summary of the statistical trend models for quantifying changes in the GEV distribution.}\label{trend_models}
\begin{center}
\begin{tabular}{|p{0.1\linewidth}||p{0.16\linewidth}|p{0.16\linewidth}|p{0.16\linewidth}|p{0.11\linewidth}|}
\hline
\textbf{Model label} & \textbf{Location trend} & \textbf{Scale trend} & \textbf{Shape trend} & \textbf{No. of parameters} ($N_\text{par}$) \\
\hline \hline
M0 & Linear in time & Constant & Constant & 4 \\ \hline
M1 & Quadratic in time &  Constant & Constant  & 5\\ \hline
M2 & Linear in time & Linear in time & Constant & 5\\ \hline
M3 & Linear in time & Constant &Linear in time  &5\\ \hline
M4 & Linear in time & Linear in time & Linear in time &6\\ \hline
\end{tabular}
\end{center}
\end{table}

\begin{table}[t]
\caption{Comparison of models in Table B3 using the Akaike information criterion (AIC).  This AIC can be thought of as quantifying the predictive skill of the various models, since we use smoothed coefficient estimates to ``predict'' the original seasonal maxima. Formally, we calculate the log-likelihood of the original seasonal maxima using the spatially-smoothed GEV coefficients for each weather station, and then averaged this log-likelihood  over all stations for M0-M4 (denoted $\overline{\log L}$). The predictive AIC is then calculated as $AIC = -2\overline{\log L} + 2N_\text{par}$,
where $N_\text{par}$ is the number of statistical parameters in each model. Smaller AIC values indicate a better statistical model; best model is in bold.}\label{AIC}
\begin{center}
\begin{tabular}{|p{0.1\linewidth}||p{0.12\linewidth}|p{0.12\linewidth}|p{0.12\linewidth}|p{0.12\linewidth}|p{0.12\linewidth}|}
\hline
\textbf{Season} & \textbf{M0} & \textbf{M1} & \textbf{M2} & \textbf{M3} & \textbf{M4}  \\
\hline \hline
DJF &\textbf{465.18} &467.41& 466.53 & 467.0881 &468.64 \\ \hline
MAM &\textbf{499.18} &501.59 &500.38 & 500.8088 &502.43 \\ \hline
JJA &\textbf{522.08} &529.5582& 523.43 & 524.11 &528.83 \\ \hline
SON &\textbf{522.97} &525.07& 524.39 &524.83 &526.30 \\ \hline
\end{tabular}
\end{center}
\end{table}


\begin{figure}[!t]
\begin{center}
\includegraphics[trim={0 0 0 0mm}, clip, width = \textwidth]{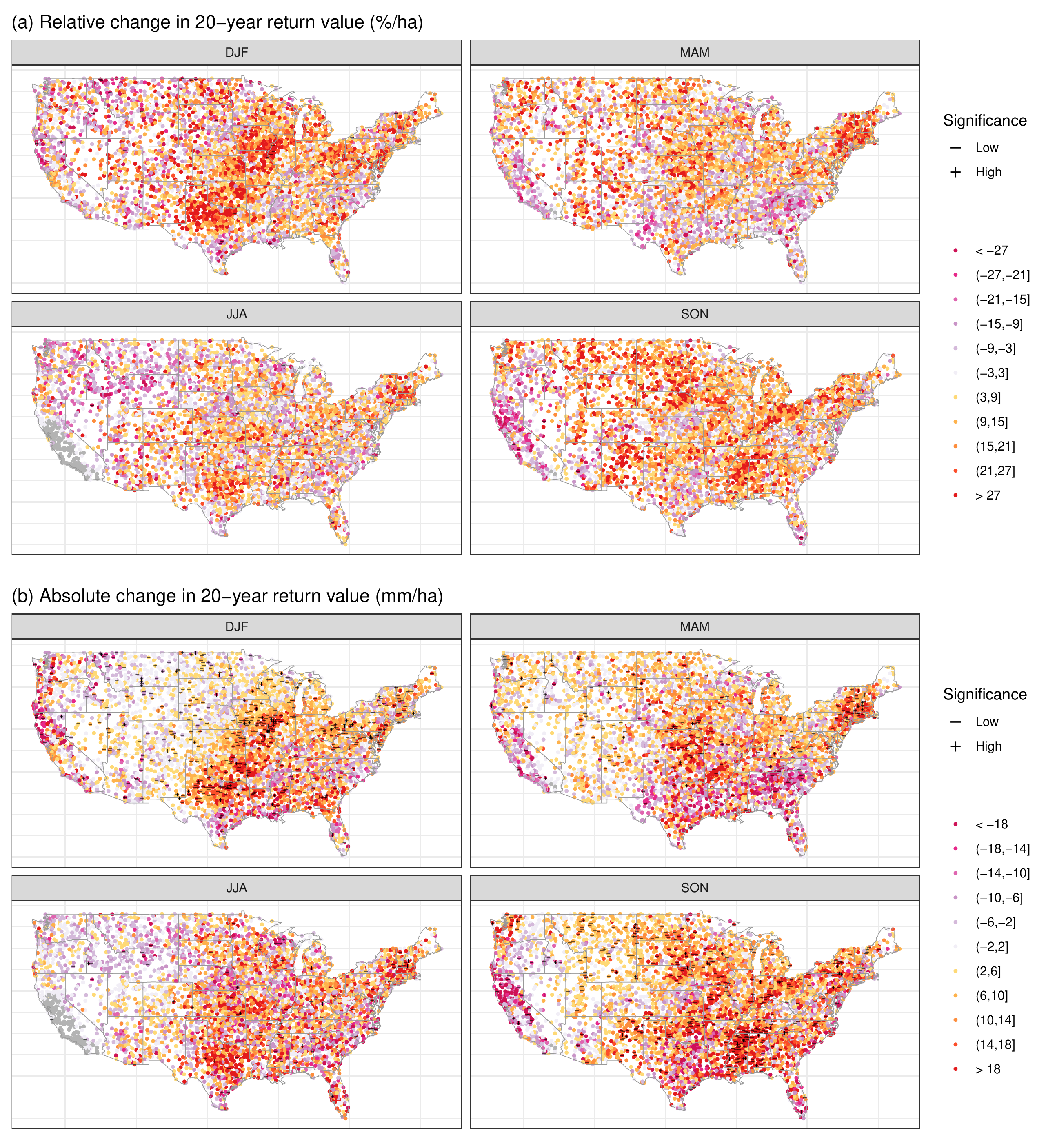}
\caption{Best estimates of the (a) relative change and (b) absolute change in the 20-year return value (1950 vs. 2017) at the station locations \textbf{without} spatial smoothing. Station locations hashed with a ``$-$'' contain a change with low statistical significance (i.e., the proportion of type I errors for these grid cells is $\leq 33\%$); those with a ``$+$'' have high statistical significance (i.e., the proportion of type I errors for these grid cells is $\leq 10\%$).}
\label{figure2_indep}
\end{center}
\end{figure}

\begin{figure}[!t]
\begin{center}
\includegraphics[trim={0 0 0 0mm}, clip, width = \textwidth]{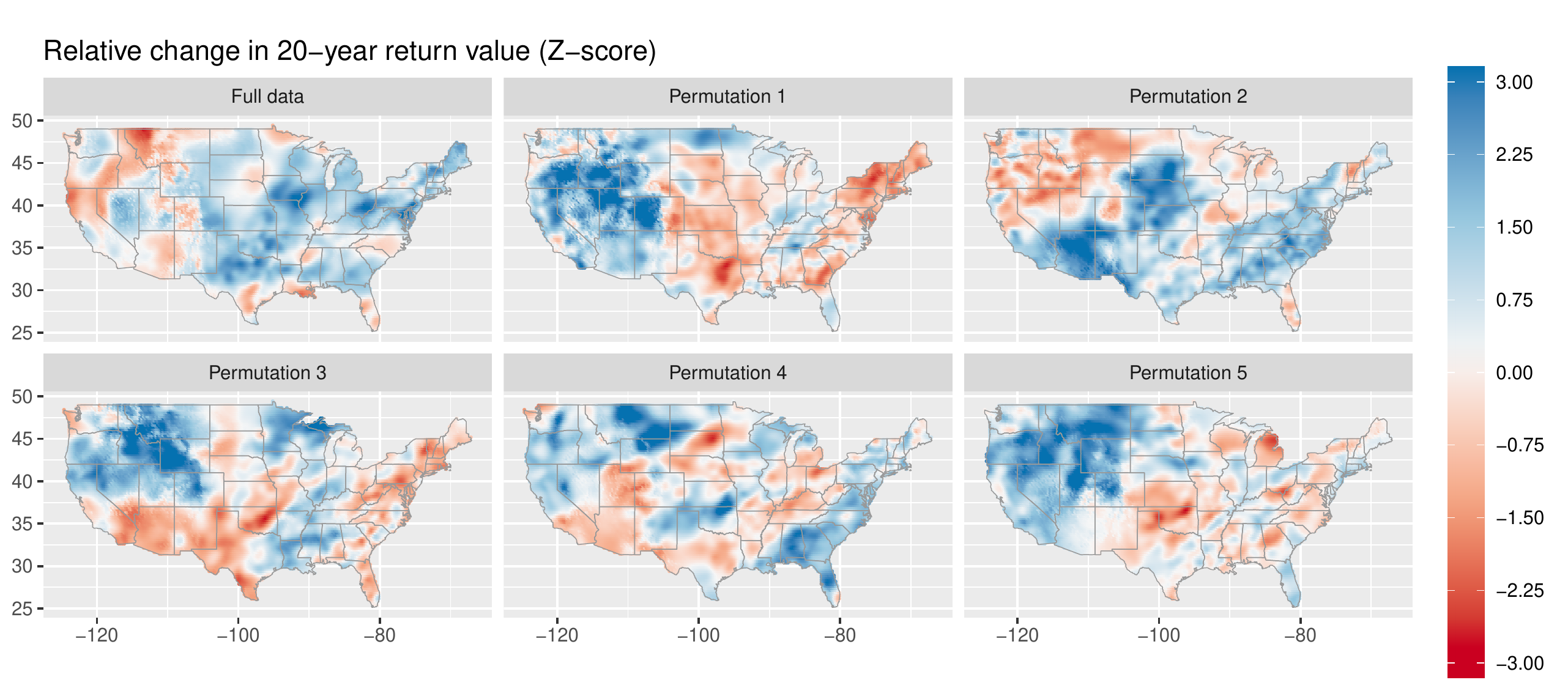}
\caption{Gridded best estimates of the standardized relative change in the 20-year return value (1950 vs. 2017) for the full unshuffled data (top left) and five selected permutation data sets.}
\label{supp_figure3}
\end{center}
\end{figure}

\begin{figure}[!t]
\begin{center}
\includegraphics[trim={0 0 0 0mm}, clip, width = \textwidth]{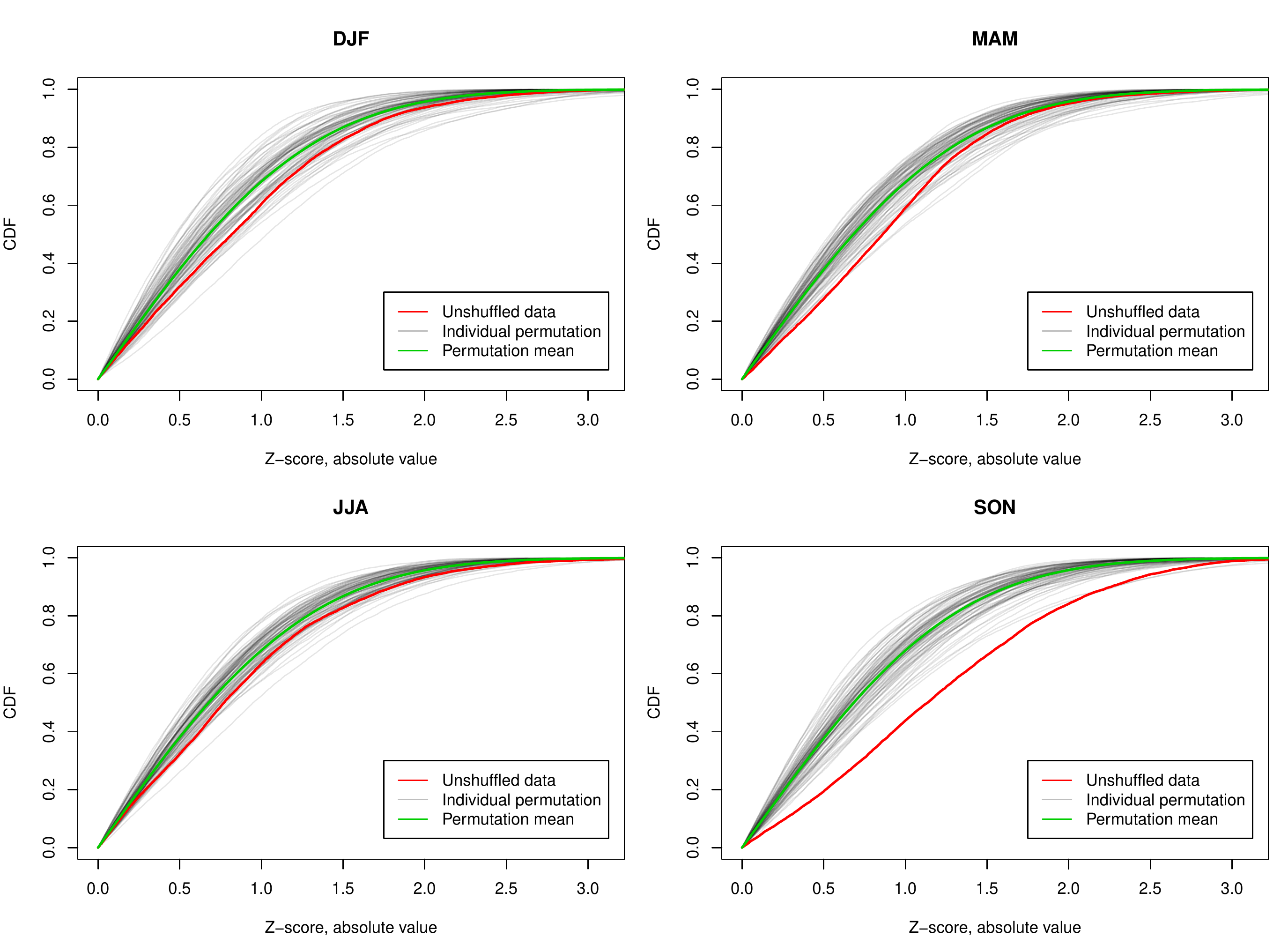}
\caption{Cumulative distribution functions (CDFs) for the relative change in return value (absolute value of the z-score) for each season. The red line refers to the CDF for the original (unshuffled) data, while the light gray lines refer to the individual permutation CDFs and their mean (in green). The unshuffled CDF lies well within the range of the permutation CDFs except for SON.}
\label{supp_figure4}
\end{center}
\end{figure}

\begin{figure}[!t]
\begin{center}
\includegraphics[trim={0 0 0 0mm}, clip, width = \textwidth]{zscore_relChange_CDF.pdf}
\caption{Cumulative distribution functions (CDFs) for the absolute change in return value (absolute value of the z-score) for each season. The red line refers to the CDF for the original (unshuffled) data, while the light gray lines refer to the individual permutation CDFs and their mean (in green). As with the relative change, the unshuffled CDF for the absolute change lies well within the range of the permutation CDFs except for SON.}
\label{supp_figure5}
\end{center}
\end{figure}

\begin{figure}[!t]
\begin{center}
\includegraphics[trim={0 0 0 0mm}, clip, width = \textwidth]{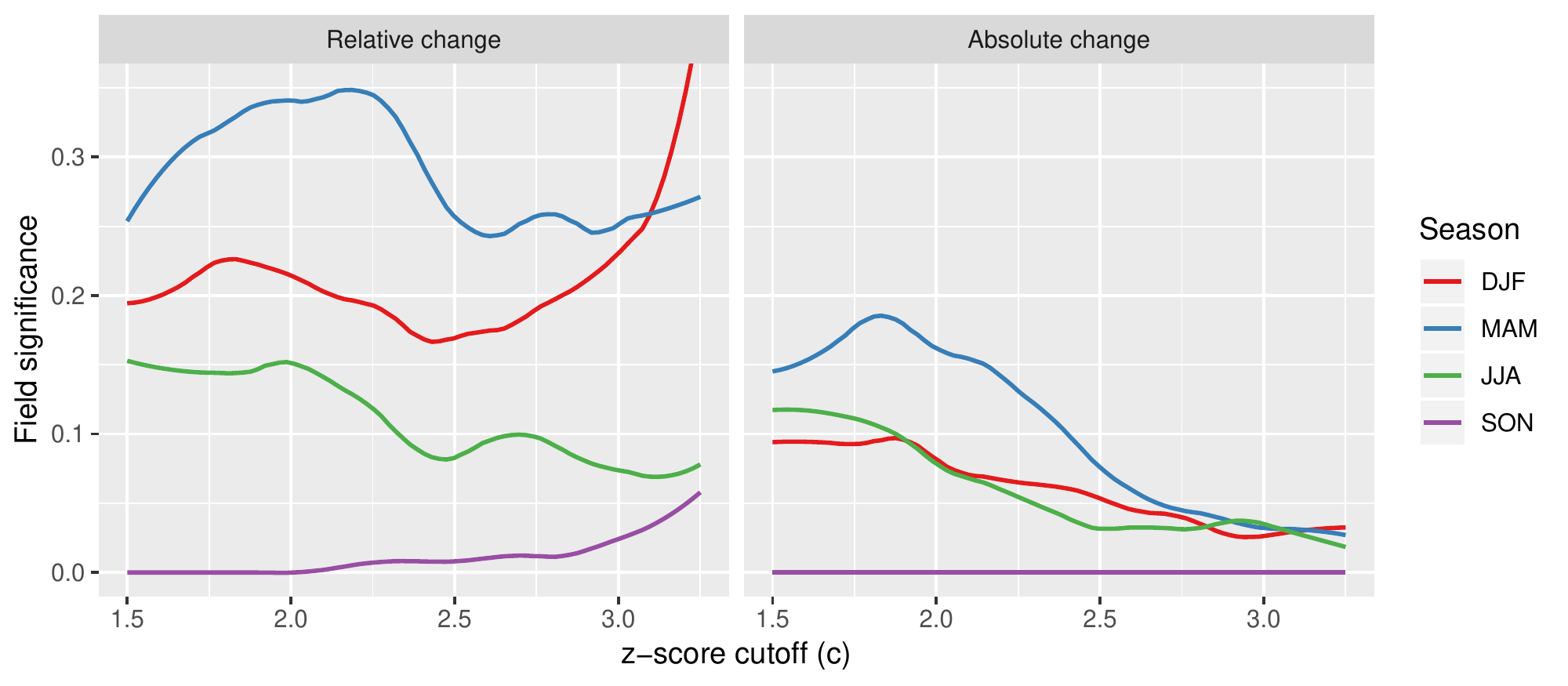}
\caption{Field significance for the seasonal changes (left: relative change in 20-year return value; right: absolute change in 20-year return value) for each of a range of $z$-score cutoffs. For the relative change, SON has a strong signal emerging for essentially all $z$-score cutoffs; otherwise, only JJA has a slightly lesser degree of field significance. For the absolute change, SON again has the strongest signal, but there appears to be an emerging signal in the other seasons as well.}
\label{supp_figure6}
\end{center}
\end{figure}

\begin{figure}[!t]
\begin{center}
\includegraphics[trim={0 0 0 0mm}, clip, width = \textwidth]{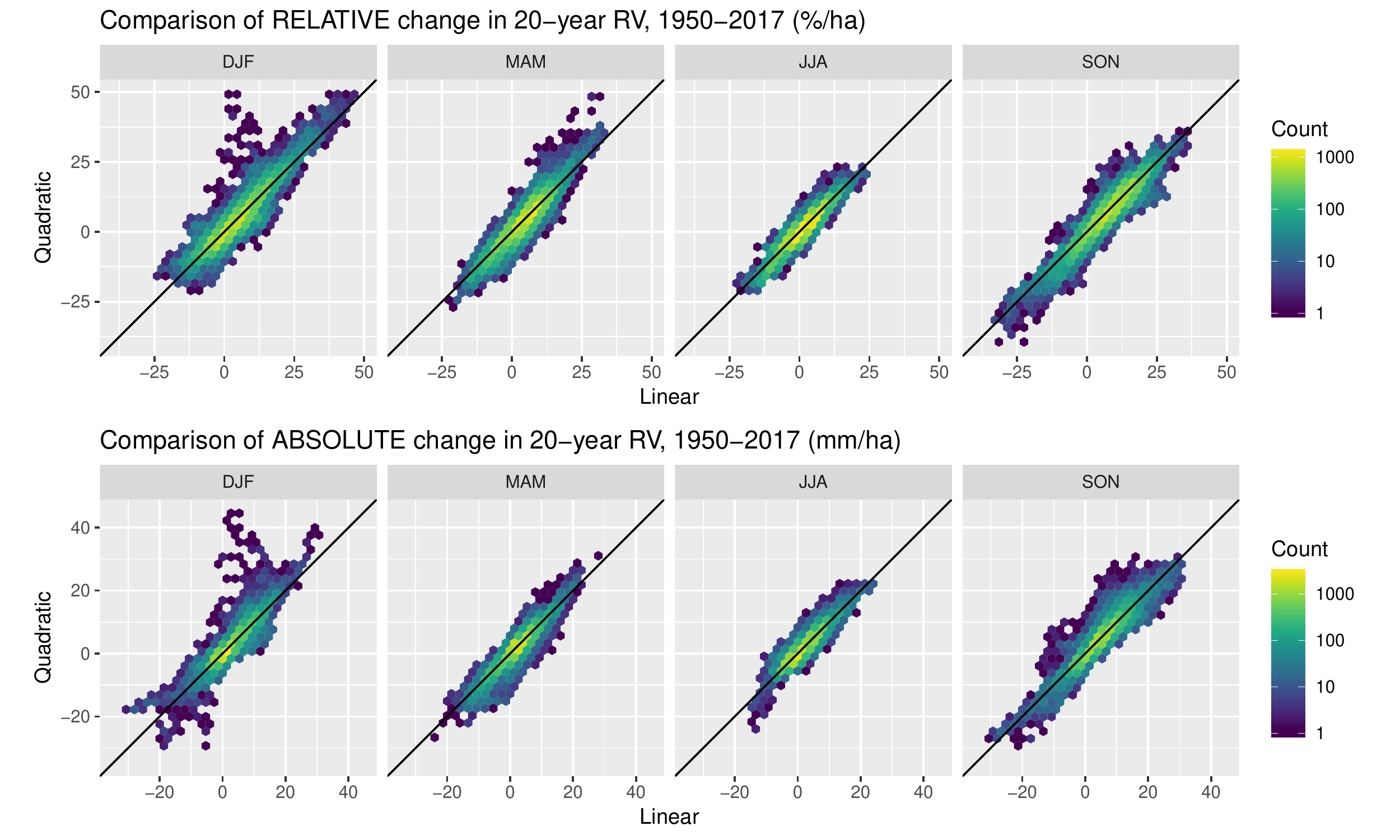}
\caption{Comparison of relative and absolute changes using a linear trend in the location parameter ($x$-axis) versus the corresponding change using a quadratic trend in the location parameter ($y$-axis). Scale and shape parameters are time-invariant for both analyses.}
\label{compareTrends} 
\end{center}
\end{figure}

\clearpage

\section{Sample size considerations for asymptotic theory}

As mentioned in the Discussion section, the Generalized Extreme Value (GEV) distribution used throughout this paper is the asymptotic (limiting) distribution of the sample maximum of a large sample of $n$ independent draws from a parent population; this theoretical result applies exactly as the sample size $n\rightarrow\infty$ \citep[see, e.g.][Chapter 3]{Coles2001}. Of course, in practice, one only has a finite sample size; furthermore, in a block maxima framework, there is a trade-off between choosing large block sizes (leading to a better approximation, i.e., less bias, but larger variance due to fewer block maxima) versus small block sizes (an increase in the bias but smaller variance). Here, we attempt to evaluate the quality of the GEV approximation for finite $n$.

A standard choice when dealing with climate data is to choose block sizes of one year; in this paper, we have chosen to apply the GEV distribution to seasonal maxima, where $n=90$ (in which case we have $T = 68$ blocks or years of data).
Two considerations are important when analyzing the extreme values of daily precipitation:
\begin{enumerate}
\item In some locations over CONUS and in some seasons, there are a large number of days with zero precipitation. Technically, the GEV approximation applies to draws from any underlying parent population (e.g., a distribution with a point mass at zero and a long right tail), but the point remains that if we are considering the ``precipitation process'' (i.e., days with nonzero rainfall), we may have fewer than 90 measurements of this process in a 90-day season.
\item Furthermore, the GEV approximation applies to a \textit{statistically independent} sample of size $n$. While daily precipitation may not exhibit the temporal autocorrelation of, e.g., temperature, the effective number of independent measurements of daily precipitation from a $90$-day season is certainly less than 90.
\end{enumerate}
In light of these considerations, we conduct a simple perfect data experiment (or simulation study) to evaluate the quality of the GEV approximation as a function of sample (block) size. For now, we set aside the spatial and temporal aspect of the analysis in this paper and focus on analyzing the extreme values of a stationary time series from a single station.

\begin{figure}[!t]
\begin{center}
\includegraphics[trim={0 0 0 0mm}, clip, width = \textwidth]{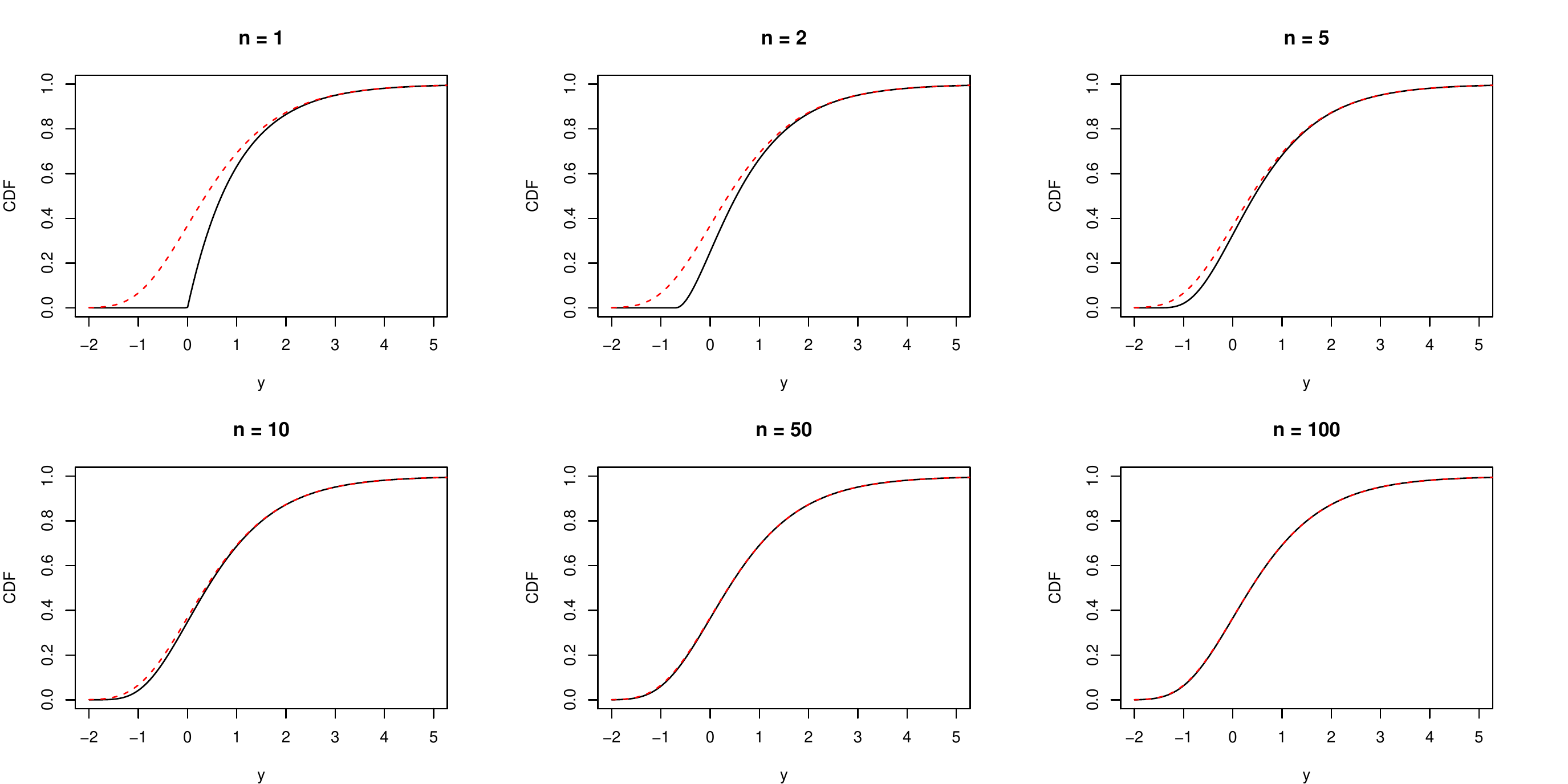}
\caption{Comparison of the cumulative distribution function (CDF) of the normalized sample maximum $F_X^n(y + \log n)$ (solid line) versus the GEV (Gumbel) limiting distribution (dashed red line) for several values of $n$. In this case, the distribution of the maximum converges quite quickly to the limiting GEV distribution. }
\label{B7} 
\end{center}
\end{figure}

For the purposes of this illustration, we generate artificial data from a known underlying distribution that has a point mass at zero and otherwise an Exponential distribution with rate 1 (the Exponential distribution with rate 1 has mean 1 and variance 1). In other words, let $\{X_1, X_2, \dots\}$ denote an independent sequence of values representing daily precipitation, where the probability distribution function of each $X_i$ is
\[
f_X(x) = p \cdot I{\{x = 0\}} + (1-p) \cdot I{\{x > 0\}} \cdot f_E(x),
\]
where $I\{\cdot\}$ is an indicator function and $f_E(x) = \exp\{-x\}$ is the Exponential distribution with rate 1. Here, $p$ is the probability of experiencing zero rainfall on a particular day. Let $F_X(x) \equiv P(X \leq x) = \int_{-\infty}^x f_X(y)dy$ denote the corresponding cumulative distribution function of $X$, and define $M_n = \max\{X_1, \dots, X_n\}$ to be the maximum value from a sample of size $n$. The cumulative distribution function of the sample maximum $M_n$ can be derived as follows:
\[
\begin{array}{rcl}
    F_{M_n}(y) \equiv P(M_n \leq y) & = & P( X_1 \leq y, \dots, X_n \leq y) \\
     & = & P( X_1 \leq y ) \times \cdots \times P(X_n \leq y) \\
     & = & F_X^n(y)
\end{array}
\]
The extremal types theorem \citep{fisher1928limiting} states that if there exist normalizing constants $a_n > 0$ and $b_n$ such that, as $n\rightarrow\infty$,
\[
P\Bigg(\frac{M_n - b_n}{a_n} \leq y \Bigg) \longrightarrow G(y),
\]
for some non-degenerate distribution $G$, then $G$ is the Generalized Extreme Value distribution. Without loss of generality, for now suppose $p=0$; if we let $a_n=1$ and $b_n = n$, it can be shown that
\[
\begin{array}{rcl}
   P\Big(\frac{M_n - b_n}{a_n} \leq y \Big) & = & F_X^n(y + \log n) \\
     & = & \Big[ 1 - \exp\{ y + \log n \} \Big]^n \\
     & = & \Big[ 1 - n^{-1}\exp\{ y \} \Big]^n \\
     & \rightarrow & \exp\big\{ -\exp\{-y\} \big\}
\end{array}
\]
as $n\rightarrow\infty$ for each $y\in \mathcal{R}$ \citep[][Chapter 3, Example 3.1]{Coles2001}. Note that $\exp\big\{ -\exp\{-y\} \big\}$ is the GEV distribution with shape parameter $\xi = 0$ (i.e., the Gumbel distribution). In this simulated example, given that we know $F_X$, the sequences $\{a_n\}$ and $\{b_n\}$, and the limiting Gumbel distribution, we can compare the approximation for various fixed values of $n$: see Figure \ref{B7}. In this case, the convergence happens quickly, with $F_X^n(y + \log n)$ and $\exp\big\{ -\exp\{-y\}\big\}$ being visually indistinguishable by the time $n = 50$. This relatively fast convergence holds even when the underlying distribution has a much heavier tail (e.g., a Pareto distribution; see \citealp{davison2015statistics} or the video here \url{http://www.annualreviews.org/doi/story/10.1146/multimedia.2015.02.23.354}). Considering the case where $p$ (the proportion of zero rainfall days) is not zero, we could simply rescale the number of measurements according to $p$ to give an ``effective'' sample size from the nonzero component of $F_X$.

\begin{figure}[!t]
\begin{center}
\includegraphics[trim={0 0 0 0mm}, clip, width = \textwidth]{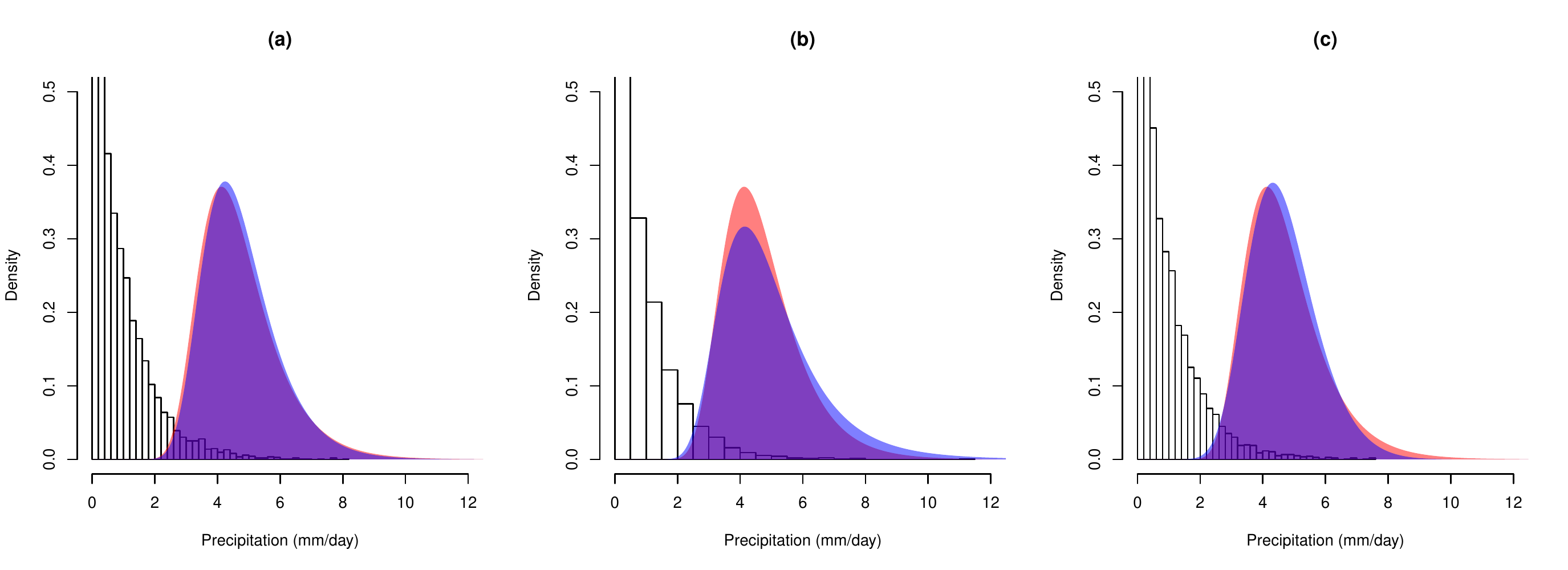}
\caption{Histograms of three simulated data sets of daily precipitation values generated from $f_X$, representing $T=68$ years of $n=90$ values with a proportion of $p=0.3$ zeros (note that the $y$-axis is truncated). The red density is the true density $f_M$ of the block (seasonal) maximum, while the blue density is the GEV approximation. In some cases the GEV approximation appears to be good (panel a), while in other cases the GEV either overestimates (panel b) or underestimates (panel c) the upper tail.}
\label{B8} 
\end{center}
\end{figure}

To consider the appropriateness of the GEV approximation more quantitatively, we now evaluate estimated return values for the simulated data. In this case where we know the form of $F_X$, we can derive exactly the distribution of the seasonal maximum:
\[
f_M(y) = n (1-p) \cdot F_E^{n (1-p) - 1}(y) \times f_E(y),
\]
where $F_E(y) = 1 - \exp\{-y\}$ is the CDF of the Exponential distribution; recall $n$ is the block size and $p$ is the proportion of zero rainfall days. The $r$-year return value is then simply the upper $1-1/r$ quantile of $f_M$. In a general case where $F_X$ is unknown (as is the case in this paper), one might use the $T$ block maximum values to estimate the $r$-year return value. As an illustration, consider Figure \ref{B8}, where we show histograms of three simulated data sets of daily precipitation values generated from $f_X$, representing $T=68$ years of $n=90$ values with a proportion of $p=0.3$ zeros. The red density is the true density $f_M$ of the seasonal maximum, while the blue density is the estimated GEV distribution for the block maxima. In some cases the GEV approximation is good (panel a), while in other cases the GEV either overestimates (panel b) or underestimates (panel c) the upper tail. These differences simply represent sampling variability in the simulated data sets.

More generally, we can explore the performance of using the GEV to estimate return values across a large number of simulated data sets. For now we maintain a fixed number of $T=68$ blocks or years, corresponding to the number of years in our actual study, but we now generate 10,000 simulated data sets for a variety of block sizes $n \in \{ 5, 10, 25, 50, 100, 200 \}$ with a fixed proportion of zero rainfall days $p=0$ (note that there is a redundancy when considering \textit{both} variable block size and variable proportion of zero rainfall days). Also, in the simulated examples, note that the data are also statistically independent. We also explore estimating the $r \in \{10, 20, 50, 100, 500, 1000\}$ year return value from these 68 blocks of data with varying block size. As in the main text, we use resampling-based techniques (the block bootstrap) to estimate the resulting standard errors in the return value estimates.

\begin{figure}[!t]
\begin{center}
\includegraphics[trim={0 0 0 0mm}, clip, width = \textwidth]{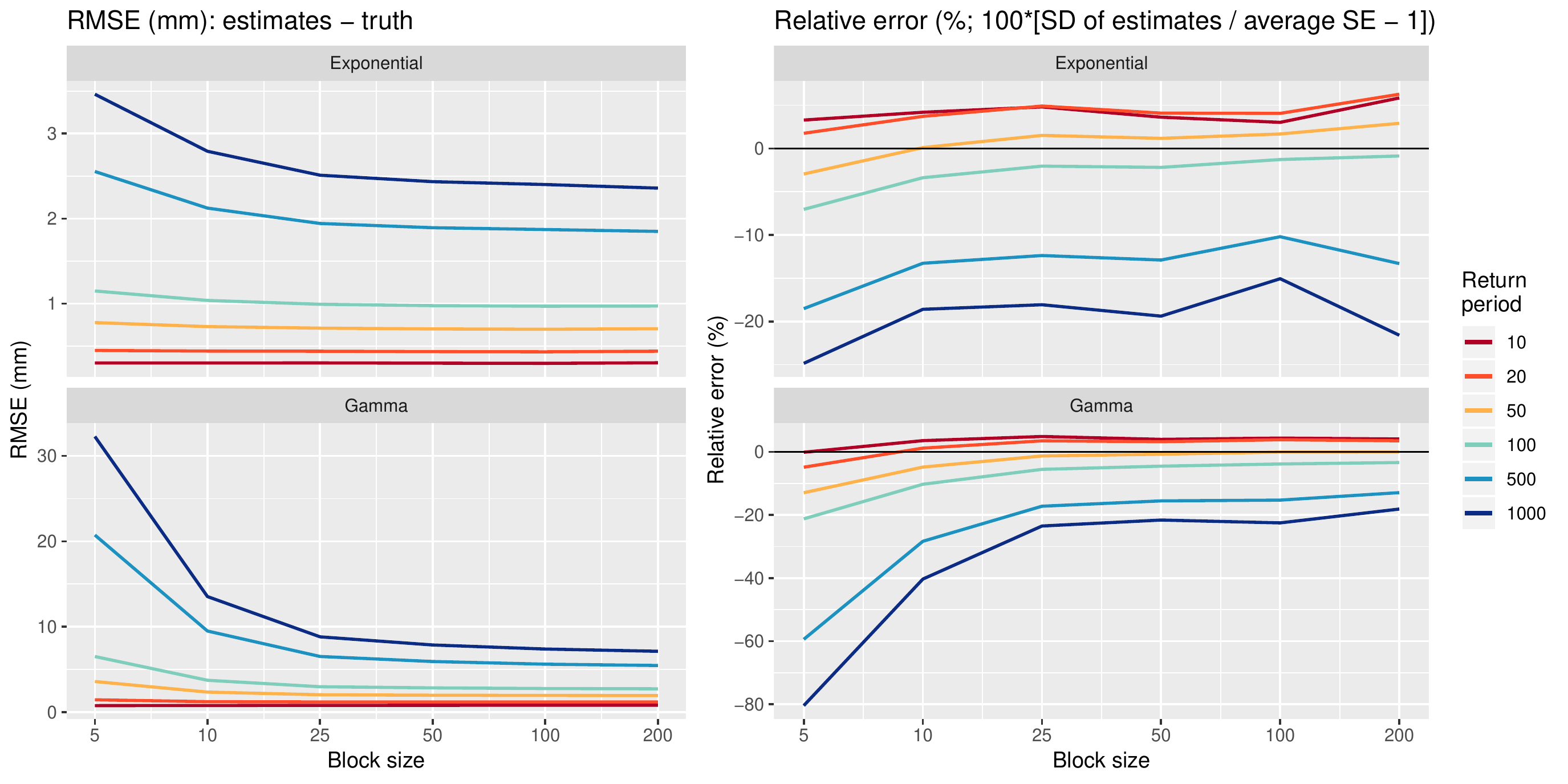}
\caption{Root mean square error (RMSE) and relative error (RE; the true Monte Carlo standard deviation divided by the average bootstrap standard error minus 1) of estimated return values, averaged over 10,000 simulated data sets from an underlying Exponential or Gamma distribution for a variety of return periods and block sizes. Smaller RMSE indicates less bias; RE greater than zero indicates standard errors that are too small (i.e., the uncertainty is anti-conservative) while RE less than zero indicates that the standard errors are too large (i.e., the uncertainty is conservative).}
\label{B9} 
\end{center}
\end{figure}

In order to evaluate the return value estimates and corresponding uncertainty quantification, we use two metrics:
\begin{enumerate}
    \item Root mean square error (RMSE): the square root of the mean squared differences (taken across the 10,000 replicates) between the true and estimated return values. RMSE evaluates the bias of the estimation procedure; smaller RMSE indicates less bias, and vice versa.
    \item Relative error (RE): the standard deviation of the return value estimates across the 10,000 replicates divided by the average bootstrap standard error of the return value estimates from the 10,000 replicates, where we subtract one from this ratio and multiply by 100 to convert to a percent error. The RE evaluates the uncertainty quantification of the estimation procedure: if the RE is zero, then the bootstrap standard errors are (on average) equal to the ``true'' (Monte Carlo) standard error, i.e., the standard errors are well calibrated. If the RE is less than 0, this indicates that the bootstrap standard errors are too large (i.e., the uncertainty is conservative); if the RE is greater than 0, this indicates that the bootstrap standard errors are too small (i.e., the uncertainty is anti-conservative).
\end{enumerate}
Figure \ref{B9} shows the results for the Exponential data previously described (the Exponential distribution with rate 1 has a mean of one and a variance of one) as well as data generated from a heavier-tailed Gamma distribution with shape $1/3$ and scale $3$ (this distribution still has mean 1 but a variance of 3). Again recall that the block sizes given here correspond to $p=0$, i.e., a zero proportion of non-rainy days, and are also statistically independent. Real daily precipitation data both (a) has zero rainfall days and (b) are not statistically independent; however, the block sizes here could be considered ``effective'' block sizes (i.e., in a 90-day season, we many only have a much smaller number of nonzero, independent measurements of precipitation).

There are several important messages from this simulation study. When estimating return values for return periods within the range of the data (up to approximately 50-year return periods), the bias of the estimates is independent of the block size: the RMSE is essentially constant across the block sizes considered. Similarly, the standard errors are well calibrated for shorter-term return periods (certainly 10- to 20- and maybe even 50-year return values) and are approximately constant across block sizes. For the 20-year return value, standard errors are on average about 4\% (2\%) too large for the Exponential (Gamma) data across the block sizes considered. This means that for shorter-term return values, the standard errors are actually slightly anti-conservative, but in any case this effect is minor. However, block size is much more important for longer-term return periods: considering the estimated 1000-year return values, the bias decays monotonically with increasing block size and is an order of magnitude larger for a block size of five versus a block size of 200. Similarly, the relative error improves monotonically with block size, such that standard errors are much too large (i.e., conservative) for small block sizes relative to the larger block sizes. These results hold for both the Exponential and heavier-tailed Gamma data, although the relative bias and lack of calibration are worse for the heavier-tailed data.

In summary, there is a non-negligible effect of block size on the quality of the GEV approximation and resulting bias/uncertainty quantification for estimating return values, but when estimating return values for return periods that are not too extreme (within the range of the data) this effect is minimal. Certainly for the 20-year return values estimated in this paper, we can be confident that our estimates contain minimal bias and the standard errors are well-calibrated. This result holds even when the effective number of measurements of daily precipitation we have from each season (due to either zero rainfall days, autocorrelation, or missing data) is as low as 25 or even 10. Of course, real daily precipitation data likely does not perfectly follow either an Exponential or Gamma distribution, but these results provide an assurance that the GEV approximation is a safe one to make for the return periods considered in the paper.

\end{appendix}

\end{document}